\documentclass[lettersize,journal]{IEEEtran}
\usepackage{amsmath,amsfonts}
\usepackage{algorithmic}
\usepackage{algorithm}
\usepackage{array}
\usepackage[caption=false,font=normalsize,labelfont=sf,textfont=sf]{subfig}
\usepackage{textcomp}
\usepackage{stfloats}
\usepackage{url}
\usepackage{verbatim}
\usepackage{graphicx}
\usepackage{cite}

\usepackage{xcolor}
\usepackage{tabularx}

\begin{document}

\title{Building time-surfaces by exploiting the complex volatility of an ECRAM memristor}


\author{Marco Rasetto, Qingzhou Wan, Himanshu Akolkar, Feng Xiong, Bertram Shi, Ryad Benosman}


\IEEEpubid{0000--0000/00\$00.00~\copyright~2023 IEEE}

\maketitle

\begin{abstract}
Memristors have emerged as a promising technology for efficient neuromorphic architectures owing to their ability to act as programmable synapses, combining processing and memory into a single device. Although they are most commonly used for static encoding of synaptic weights, recent work has begun to investigate the use of their dynamical properties, such as Short Term Plasticity (STP), to integrate events over time in event-based architectures. However, we are still far from completely understanding the range of possible behaviors and how they might be exploited in neuromorphic computation. This work focuses on a newly developed Li$_\textbf{x}$WO$_\textbf{3}$-based three-terminal memristor that exhibits tunable STP and a conductance response modeled by a double exponential decay. We derive a stochastic model of the device from experimental data and investigate how device stochasticity, STP, and the double exponential decay affect accuracy in a hierarchy of time-surfaces (HOTS) architecture. We found that the device's stochasticity does not affect accuracy, that STP can reduce the effect of salt and pepper noise in signals from event-based sensors, and that the double exponential decay improves accuracy by integrating temporal information over multiple time scales. Our approach can be generalized to study other memristive devices to build a better understanding of how control over temporal dynamics can enable neuromorphic engineers to fine-tune devices and architectures to fit their problems at hand.
\end{abstract}


\section{Introduction}
\IEEEPARstart{T}{he} last decade has brought considerable progress in AI, mainly owing to the advent of Graphics Processing Units (GPUs) and other hardware accelerators. However, this progress has not been matched from the perspective of emulating general intelligence and cognition. Ideas such as deep multilayer learning and backpropagation have helped solve a particular class of well-defined problems but require high energy and vast amounts of labeled data \cite{Pierson2017,datascaling_chinchilla}. These requirements drastically limit on-board ''intelligence'' and reduce autonomy. Thus, essential functionalities such as continuous always-on learning with a reasonable power budget are still out of reach. 

Neuromorphic engineering holds the promise of mitigating these restrictions \cite{schuman2017survey,lichtsteiner2006128}. Recently, this field has reached a level of maturity that allows it to impact several other domains where autonomy and low power on-the-edge computation are crucial. One core principle is to remove the separation of memory and computation, typical of Von Neumann architectures, by taking inspiration from neurons and synapses and more closely integrating computation and memory.

Many researchers have taken an interest in memristive devices, given their ability to implement tunable nonvolatile weights similar to synaptic efficacy in biological synapses. Following this paradigm, memristors have been applied to many network-based computational approaches. Doygu et al. simulated memristor networks able to learn sequences of inputs \cite{2011SequenceMemristor}. Suri et al. used another simulated network of phase-changing memristors to learn MNIST letters\cite{2011PCMComplexVisualPatterns}. Systems that perform STDP (Spike Time Dependent Plasticity) using RRAM (REsistive RAM) devices \cite{Ambrogio_2013_RRAM} or that implement recursive networks using PCM memristors have also been reported \cite{Recursive2014PCM}. Vincent et al. simulated a network of STT-MRAM devices for car detection \cite{2015STT-MRAMNetwork}. Networks of memristors have also been proposed to implement the k-means algorithm \cite{Kmeans2018} and unsupervised learning \cite{wang2018fully}. In all of these examples, the synaptic devices modeled using memristors working in a ``static'' fashion, i.e., as a fixed scalar multiplier of spike events, before integration by the ``neuron membrane''. This approach has been widely adopted because of its simplicity of implementation and mathematical tractability, as synaptic operations can be described by simple linear algebra.

In recent years, a different approach has surfaced. Several works have demonstrated the presence of transient conductance responses in memristive devices akin to short-term plasticity (STP) and Excitatory/Inhibitory Post Synaptic Potentials (EPSP/IPSP)\cite{Xia2019,wan2019emerging,yang2013filRRAM,ohno2011stpCBRAM}. These memristors with ``volatile'' properties are extremely promising for implementing neuromorphic networks that need physical devices capable of temporal computation \cite{berdan2016emulating}.
In these devices, input events (i.e., voltage pulses) cause temporary changes in conductance that exponentially relax back to baseline, like EPSPs and IPSPs in biological synapses. Moreover, the conductance change is potentiated when multiple input events are close in time (STP). These short-term dynamics allow the modeling of temporal kernels and short-term plasticity without the need for additional circuitry. Electrochemical memristors exhibit both STP and EPSPs simultaneously \cite{sharbati2018Elect,double_decay_mem}. However, in memristors with oscillatory properties \cite{Hua2019,Lim2016,Pickett2013}, STP is not present. Moreover, some devices can produce EPSPs with multiple exponential decays, making the range of possible dynamics extremely complex. We are still far from understanding the entire repertoire of short-term dynamics in volatile memristors, let alone being able to exploit them in real, practical scenarios. Many works utilizing the short-term dynamics of memristive devices have demonstrated only simple networks (or even single neurons) operating on limited examples\cite{Sung2022,Sarwat2022,Wan_et_moi,sun20182D,berdan2016emulating,yang2013filRRAM,ohno2011stpCBRAM}, or use memristor dynamics for different spiking neuron operations such as adaptive thresholds \cite{shaban2021adaptive} and synaptic traces \cite{demiraug2021pcm}. To the best of our knowledge, the relationship between the different types of dynamics in representing temporal information from spiking data and recognition rates on real-world neuromorphic datasets has not been studied. Understanding this relationship will enable the design of custom neuromorphic systems that make full use of memristive dynamics for efficient computation.

This paper studies the effect of different memristive dynamics on recognition accuracy by complex networks on complex tasks by simulating a neuromorphic architecture based on a LiWES artificial synapse: a three-terminal electrochemical memristor with double exponential decays in the order of ten to hundred milliseconds (required by many neuromorphic datasets) and STP \cite{Wan_et_moi}. 
We build a model of the device response and use it to simulate a network based on the Hierarchy of Time Surfaces (HOTS) architecture \cite{lagorce2016hots} for pattern recognition. While these architectures tend to have lower accuracy than SNNs trained with backpropagation, the local unsupervised learning used for feature extraction is well suited for on-chip learning, especially in conjunction with memristive technologies \cite{Kmeans2018}. 

We evaluate performance on the N-MNIST \cite{NMNIST} and POKERDVS \cite{POKERDVS} datasets. First, we study whether inherent stochasticity in memristive dynamics could negatively impact accuracy. We compare the device model against an ideal noiseless memristor, finding no significant difference in the classification accuracy. 
Using information theoretic measures, we also study the effect of stochasticity on the mutual information of events propagated by the network and discuss a possible solution to reduce accuracy loss in architectures using dynamics in noisy memristors. Finally, we break down the model to see how different memristive dynamics (multiple-exponential decays and STP) can be used to improve accuracy and performance in HOTS-like neuromorphic architectures.

\section{The Li$_\textbf{x}$WO$_\textbf{3}$ electrochemical memristor}

``Volatile'' memristors enable efficient implementations of temporal computing, as they combine temporal dynamics and short-term plasticity in a single device. Among these, electrochemical memristors~\cite{FirstLi} have become good candidates thanks to their low power consumption, linear and symmetric response, low variability, and high reliability \cite{FirstLi,li2019low,fuller2019parallel,yang2018all,double_decay_mem,qian2016artificial,Wan_dynamics,van2017non,yao2020protonic,li2021one,li2020filament}. 
We have previously proposed the use of a novel electrochemical memristor~\cite{Wan_et_moi} based on Lithium Ions and Tungsten Oxide (Li$_\textbf{x}$WO$_\textbf{3}$), which has the advantages of low programming voltage (0.2 V), fast programming speed (500 ns), and high precision (1024 states corresponding to 1024 10ms 0.5V ``write'' pulses before reaching saturation), wide conductance range ($\sim 1\mu$s to $\sim 200\mu$s), with a channel area of 400x200$\mu m^2$.
These devices have been used to model synapses and to implement electrochemical random access memory (ECRAM) \cite{tang2018ecram}. They are especially suitable for neuromorphic networks because they can model synaptic dynamics and short-term plasticity (STP) with time constants ranging from a few to hundreds of milliseconds. We focus on a version of Li$_\textbf{x}$WO$_\textbf{3}$ memristor that uses a self-gate design in which transitory effects dominate long-term effects\cite{Wan_dynamics}.

\begin{figure}[ht]
\centering
    \includegraphics[scale=0.24]{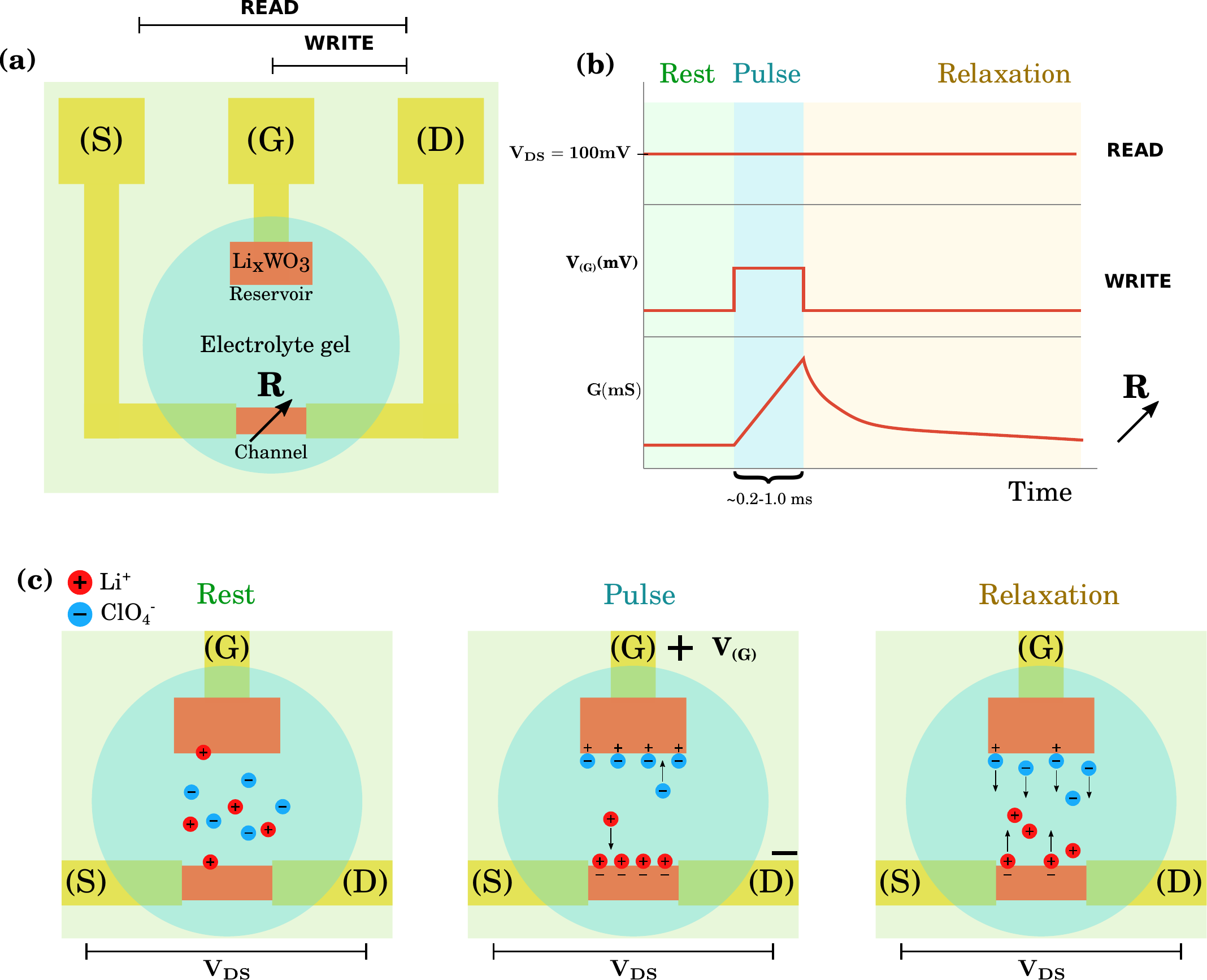}
    \caption{The structure and operation of the physical  Li$_\textbf{x}$WO$_\textbf{3}$ electrochemical synapse modelled in this paper. (a) The Li$_\textbf{x}$WO$_\textbf{3}$ electrochemical synapse is a three-terminal device with a (S)ource, (G)ate and (D)rain. The gate and channel between S and D are built by deposition of tungsten oxide (WO$_\textbf{3}$) films (red) on a LaALO$_\textbf{3}$ substrate (green). The films are connected to gold terminals (yellow). Lithium ($\mathrm{Li+}$) and $\mathrm{ClO_4-}$ ions are introduced by applying a drop of electrolyte gel (blue) on top of the device. (b) The electrical behavior of the memristor in response to a square WRITE voltage pulse applied between (G) and (D) and a small DC READ voltage (0.1 V) applied between (S) and (D). (c) The electrochemical behavior of the memristor.  
    }
    \label{fig:WO3 device}
\end{figure}

This structure of the Li$_\textbf{x}$WO$_\textbf{3}$ memristor is illustrated in Fig. \ref{fig:WO3 device}(a). Unlike conventional two-terminal memristors, the Li$_\textbf{x}$WO$_\textbf{3}$ electrochemical is composed of three terminals, the (S)ource, (D)rain and (G)ate. The memristor is built by deposition of tungsten oxide (WO$_\textbf{3}$) films on a LaALO$_\textbf{3}$ substrate. Lithium ions introduced via an electrolyte gel can flow between the gate and channel. When embedded into the tungsten oxide films, they act as short or long-term doping charges, changing the film conductance \cite{Wan_et_moi}. The conductance between the source and the drain terminal, $G_\mathrm{DS}$, is considered to be the synaptic weight of the device.

The memristor's electrical behavior is illustrated in Fig. \ref{fig:WO3 device}(b). The conductance $G_\mathrm{DS}$ can be read by applying a small DC reading voltage ($V_\mathrm{DS}$ = 100mV) across the source and drain. During the ``write'' phase, we modulate $G_\mathrm{DS}$ by applying a square voltage pulse between the gate and drain. The channel conductance increases linearly. Once the pulse is removed, the conductance decays following a double exponential decay function. Reading can be carried out in parallel and does not interfere with writing. This design allows for lower programming voltages and better state retention~\cite{van2017non}. State retention and low programming voltage are important when designing standard memristive networks for spiking and artificial neural network implementations. These allow for stable networks with low power consumption. Multiple small pulses can be used to change the device conductance gradually, allowing the artificial networks to mimic their biological counterparts. 

Fig. \ref{fig:WO3 device}(c) illustrates the electrochemical operation. At rest (no gate to drain voltage applied), the Li+ and $\mathrm{ClO}_4-$ ions are in equilibrium between the tungsten oxide films and the electrolyte gel.  The positive ``write'' pulse creates an electric field, which causes Li+ ions to accumulate at the channel/electrolyte interface and charge-balancing $\mathrm{ClO}_4-$ ions to accumulate at the gate. Doping of the channel by the Li+ ions increases the channel conductance. During the relaxation phase, after the removal of the write pulse, the ions return to equilibrium and the channel conductance returns to its resting value. The double exponential decay response can be explained by the Kohlrausch-Williams-Watts (KWW) relaxation model \cite{williams1970non, li2017electric}, which has also been found in other electrochemical devices \cite{double_decay_mem}.

Changing the material properties enables ``programming'' the exponential decays in the range of tens to hundreds of milliseconds, which is optimal for temporal integration of events for many real-world datasets \cite{lagorce2016hots,purves2001neuroscience}. Due to the double exponential decay, closely spaced pulses generate accumulation. This property allows the memristor to exhibit short-term plasticity (STP) similar to biological neurons. Moreover, the high number of pulses required to reach saturation (1024) make it more than capable to work on the proposed datasets, where the maximum number of events per pixel is 48, corresponding to a max of 48 ``write'' pulses per single synapse/device. Together, these properties make this device an excellent candidate for studying the computational benefits of memristors' dynamics for neuromorphic time based learning applications.

\subsection{Mathematical model of synaptic dynamics}
\label{sec:mathmodel}

Following \cite{williams1970non,li2017electric,Wan_dynamics,Wan_et_moi,double_decay_mem}, we develop a mathematical model to predict the electrical behavior of the $\mathrm{Li}_x \mathrm{WO_3}$ memristor by first modelling its response to a single square write pulse, then combining the responses. 

Given a spike train with spike times  $\{t_i\}$ indexed by $i \in \{0,1,2,\ldots \}$, the conductance response is given by 
\begin{equation} \label{eq:G(t)}
G(t)= \sum_{i} 
\left( G_{1,i}(t)+ G_{2,i}(t) \right) 
+ \eta(t) 
\end{equation}
where $G_{1,i}(t) + G_{2,i}(t)$ is the memristor's response to all spikes up to and including the $i$th spike for times $t \in (t_i,t_{i+1}]$, and $\eta(t)$ is zero-mean Gaussian white noise with variance $\sigma^2$. To model the double exponential decay, we express the response as the sum of two components, $G_{k,i}(t)$ for $k \in \{1,2\}$, each modelling a single exponential response.  

\begin{figure}[t]
  \centering
  \includegraphics[width=\columnwidth]{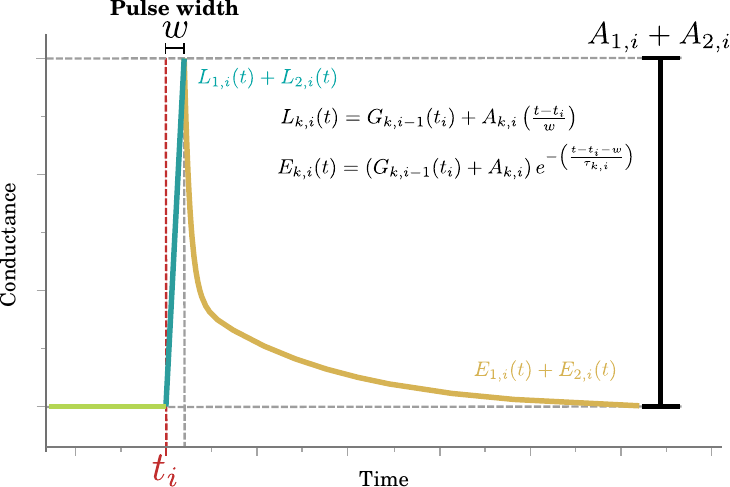}
  \caption{Example response of our model to a pulse at $t_i$ with pulse width $w$ assuming initial steady state (in green). The model presents a linear increase (in blue) for the duration of the pulse. At its end, it will relax to a steady state following a double exponential decay (in gold).} 
  \label{fig:Fitting}
\end{figure}

We define the components $G_{k,i}(t)$ recursively:
\begin{equation} \label{eq:Gki(t)}
G_{k,i}(t) = 
\begin{cases}
L_{k,i}(t) & 
\text{for } t_i < t \leq \mathrm{min} ( t_i+w, t_{i+1} ) \\
E_{k,i}(t) & \text{for $t_i+w < t \leq t_{i+1}$} \\
0 & \text{otherwise}
\end{cases}
\end{equation}
where $w>0$ is the width of the write pulse.
$L_{k,i}(t)$ models the linear rise in conductance starting from $G_{k,i-1}(t_i)$ to $G_{k,i-1}(t_i) + A_{k,i}$, where $A_{k,i}$ is the peak conductance change due to the $i$th write pulse.\begin{equation} \label{eq:Lk(t)}
L_{k,i}(t) = G_{k,i-1}(t_i) + A_{k,i}  
\left( \frac{t-t_i}{w} \right) 
\end{equation}
$E_{k,i}(t)$ models an exponential decay in conductance from $G_{k,i-1}(t_i) + A_{k,i}$ to zero with time constant $\tau_{k,i}$. 
\begin{equation} \label{eq:Ek(t)}
E_{k,i}(t)=\left( G_{k,i-1}(t_i) + A_{k,i} \right) 
e^{- \left( \frac{t-t_i-w}{\tau_{k,i}} \right)}
\end{equation}
Fig. \ref{fig:Fitting} shows our model for a given pulse width $w$ at time $t_i$. 

 We can see from (\ref{eq:Gki(t)}), (\ref{eq:Lk(t)}) and (\ref{eq:Ek(t)}) that $G_{k,i}(t)$ depends on the conductance at the start of the $i$th pulse, $G_{k,i-1}(t_i)$. This models STP, where past pulses all contribute to the current device conductance. To model the hypothetical memristor without STP in section IV C, we remove  $G_{k,i}(t)$ from (\ref{eq:Lk(t)}) and (\ref{eq:Ek(t)})
\begin{equation} \label{eq:Lk(t)-noSTP}
L^{\overline{\mathrm{stp}}}_{k,i}(t) = A_{k,i}  
\left( \frac{t-t_i}{w} \right) 
\end{equation}
\begin{equation} \label{eq:Ek(t)-noSTP}
E^{\overline{\mathrm{stp}}}_{k,i}(t)= A_{k,i}
e^{- \left( \frac{t-t_i-w}{\tau_{k,i}} \right)}
\end{equation}
so that every new pulse resets the peak conductance to $A_{k,i}$, rewriting any conductance value from previous pulses.

As we are also interested in studying the effect of device stochasticity on computation, we consider two models: an ideal model and a stochastic model. 

In the ideal model, the peak conductance changes and time constants are the same for all pulses, i.e., $A_{k,i} = \overline{A_k}$ and $\tau_{k,i} = \overline{\tau_k}$ for $k \in \{1,2\}$, where $\overline{A_k}$ and $\overline{\tau_k}$ are positive constants. The noise is zero $(\eta(t)=0)$. 

In the stochastic model, the $A_{k,i}$ and $\tau_{k,i}$ are drawn from an independent and identically distributed discrete time (i.i.d.) random process in $i$, where each sample is drawn from a Gaussian distribution,
\begin{equation} \label{eq:Adist}
A_{k,i} \sim \mathcal{N}(\overline{A_k},\sigma_{A_k}^2).
\end{equation}
\begin{equation} \label{eq:taudist}
\tau_{k,i} \sim 
\mathcal{N}(\overline{\tau_k},\sigma_{\tau_k}^2).
\end{equation}
where negative samples are rectified. 

The dynamics of the LiWES can be tuned by changing the write pulse properties, such as pulse width and amplitude \cite{Wan_et_moi}.
To replicate the device behavior for a given pulse, we need to obtain the mean and standard deviation of the model parameters ($A_1,A_2,\tau_1,\tau_2$). We first compute the rest conductance by averaging the conductance before pulse onset and subtracting this from the data. We then combine (\ref{eq:G(t)}) and (\ref{eq:Gki(t)}) to compute the response, which we split into two parts:
\begin{equation}
G_\mathrm{rise}(t) = (A_1 + A_2) \left( \frac{t-t_0}{w} \right) \text{  for } t_0 < t \leq t_0+w
\end{equation}

\begin{equation}
G_\mathrm{decay}(t) = A_1 e^{- \left( \frac{t-t_0-w}{\tau_{1}} \right)}
+ A_2 e^{- \left( \frac{t-t_0-w}{\tau_{2}} \right)} \text{  for } t_0+w < t
\end{equation}
We fit the model parameters ($A_1,A_2,\tau_1,\tau_2$) using the least squares fit between the experimental data after the write pulse ($t > t_0+w$) and $G_{decay}(t)$ using the Gauss-Newton algorithm. We use the model and fitted parameters to predict the responses for $t \leq t_0+w$. 
  
To find the parameters of the stochastic model, we repeated the model fit using data from multiple recordings of pulses with different pulse amplitudes (ranging from 1$V$ to 4$V$) and durations (ranging from 200$\mu$s to 1~ms). Based on these fits, we calculated the distributions of the model parameters. We report the fitting results in Table \ref{tab:FixedV} and Table \ref{tab:FixedT}. The tables show the mean and standard deviations of the parameter estimates fitted over 20 recordings for the 200us 1V pulse and 5 recordings for the remaining conditions. The variance of the noise $\sigma_\eta$ was set to the mean square error of the fit.
For details, refer to the supplementary materials. 

\begin{table}[ht] 
\centering
\caption{Model parameters (mean $\pm$ s.d.) obtained with a pulse with fixed 1V amplitude but varying duration.}
\begin{tabularx}{\columnwidth}{l|lllll}
1V & ${A_1}$      & ${\tau_1}$      & ${A_2}$      & ${\tau_2}$        & $\sigma_\eta$ \\ \hline
200us & 0.57$\pm$.27 & 5ms$\pm$2ms  & 0.5$\pm$.05  & 92ms$\pm$18ms  & 0.0464   \\
500us & 0.74$\pm$.18 & 16ms$\pm$9ms & 0.26$\pm$.06 & 588ms$\pm$31ms & 0.0245   \\
750us & 0.78$\pm$.19 & 10ms$\pm$2ms & 0.23$\pm$.03 & 513ms$\pm$98ms & 0.0149   \\
1ms   & 0.75$\pm$.29 & 10ms$\pm$3ms & 0.22$\pm$.02 & 390ms$\pm$68ms & 0.0097   \\ \hline
\end{tabularx}
\label{tab:FixedV}
\end{table}

\begin{table}[ht]
\centering
\caption{Model parameters (mean $\pm$ s.d.) obtained with a pulse with fixed 200us duration but varying amplitude.}
\begin{tabularx}{\columnwidth}{l|lllll}
200us & ${A_1}$      & ${\tau_1}$      & ${A_2}$      & ${\tau_2}$        & $\sigma_\eta$ \\ \hline
1V    & 0.57$\pm$.27 & 5ms$\pm$2ms  & 0.5$\pm$.05  & 92ms$\pm$18ms   & 0.0464   \\
2V    & 0.54$\pm$.29 & 7ms$\pm$1ms  & 0.35$\pm$.02 & 122ms$\pm$20ms  & 0.0244   \\
3V    & 0.77$\pm$.24 & 13ms$\pm$6ms & 0.23$\pm$.02 & 373ms$\pm$98ms  & 0.0189   \\
4V    & 0.75$\pm$.17 & 11ms$\pm$3ms & 0.25$\pm$.01 & 501ms$\pm$101ms & 0.0159   \\ \hline
\end{tabularx}
\label{tab:FixedT}
\end{table}

\begin{figure}[ht]
  \includegraphics[width=\columnwidth]{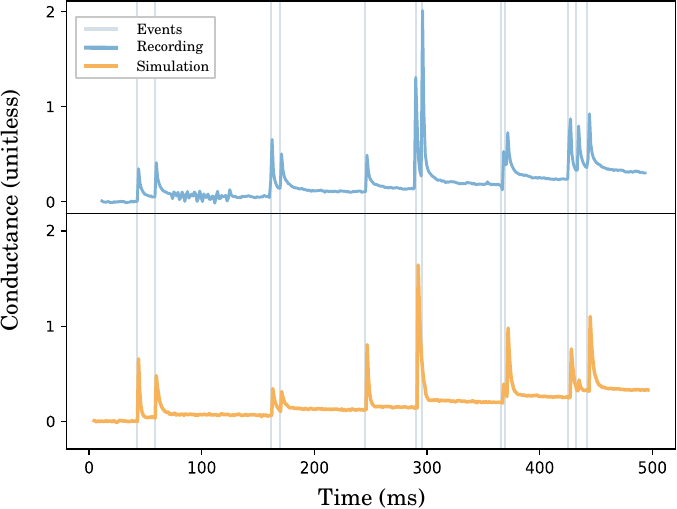}
  \caption{Our stochastic model (in orange) and a memristor recording (in blue) for a pulse of 1V 200us. We sample the response parameters from the parameter distributions obtained from fitting our model to experimental data.}
  \label{fig:Simulation_comparison}
\end{figure}

We can use equations (\ref{eq:G(t)})-(\ref{eq:taudist}) and the distributions specified in Tables \ref{tab:FixedV} and \ref{tab:FixedT} to simulate the response of any number of devices to any set of spike trains. Figure \ref{fig:Simulation_comparison} compares the output of our stochastic model with the response of an actual device to the same spike train. Note that due to the stochasticity, we do not expect spike-to-spike matching of the responses. Rather, the statistics and timing of the responses will be similar. 

\section{Memristor Implementation of HOTS}

The Hierarchy of Event-Based Time-Surfaces (HOTS) architecture is a neuromorphic architecture for unsupervised pattern recognition \cite{lagorce2016hots}. HOTS networks are highly versatile and can be applied to the output of neuromorphic sensors for different modalities \cite{lagorce2016hots,stereoHOTS,gauvain2021optogenetic,rasetto2021event}. Moreover, HOTS neurons are more mathematically tractable than other Spiking Neural Network (SNN) models. This feature makes it easier to isolate the effects of memristive dynamics on network behavior using the analysis we describe in the next section.

A HOTS network maps each input spike to an output spike from one of the neurons in the network. HOTS networks draw inspiration from clustering. Each neuron in a HOTS network represents a cluster corresponding to a pattern of events within a spatial window. The diversity of possible input patterns is reflected by the diversity of patterns represented by the neurons in the network.  In order to represent event patterns as points to cluster, HOTS introduces the concept of the Time-Surface. Every time a neuron produces an event, it triggers the creation of a Time-Surface, an array representing the time history of events at that and neighboring neurons.

\begin{figure*}
  \centering
  \includegraphics[scale=0.73]{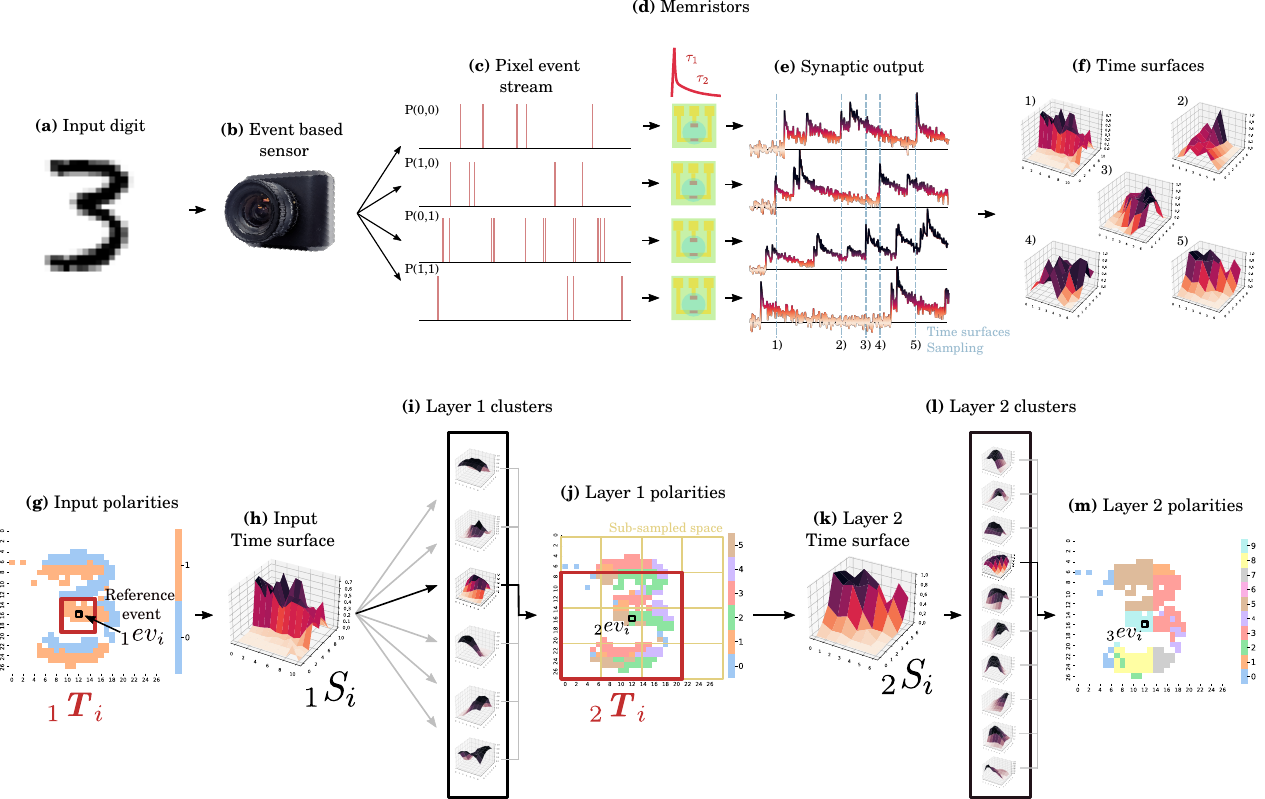}
  \caption{Top row (a-f): Mapping spatio-temporal event patterns into Time-Surface features using memristive synapses. Bottom row (g-m): N-MNIST classification using the memristive HOTS network. An input digit (a) is presented to the event-based sensor (b), which produces asynchronous event streams (c) at each pixel based on the luminance variation over time. Each pixel stream is input to a memristor (d), which interpolates the spike train with exponential decay kernels (in red) with time constants $\tau_1$ and $\tau_2$, resulting in time-varying conductances (e). For each new event $ev_i$ (indicated in light blue), we generate a Time-Surface by sampling the conductances from memristors at neighboring pixels, resulting in a 2D map that encodes temporal correlations between events at different pixels (f). The spiking activity (g) from the event-based sensor has two polarities, indicating increases (orange) or decreases (blue) in luminance. This figure shows the polarity of the last spike at each pixel, if any, during the last 10ms before a reference event $ev^1_i$. Time surfaces (h) are created by sampling changes to memristor conductances in a square neighborhood around the reference event, shown in red. Time surfaces are clustered (i). Input events produce output events (j) with the same location and time stamp, but polarity determined by the closest cluster index. This new set of events is spatially sub-sampled, resulting in a larger effective neighborhood size. In the next layer, these are used as inputs to generate new time surfaces (k) that are again mapped to clusters (l) to produce output events (m). This process can be repeated multiple times to increase the temporal and spatial scale of the classified features.
  }
  \label{fig:Architecture}
\end{figure*}

This section describes a model of a HOTS implementation that exploits the dynamics of Li$_\textbf{x}$WO$_\textbf{3}$ memristors to create the time surfaces. The process of creating time surfaces is shown in Fig. \ref{fig:Architecture}, where we assume input comes from an event-based vision sensor and the task is digit recognition. Input images (a) are captured by the event-based sensor (b), which emit trains of events at each pixel in response to brightness changes (c)\cite{lagorce2016hots}. 

Each event $i$ from an event-based vision sensor can be described by the tuple:
\begin{equation} \label{eq:event}
ev_i = (x_i,y_i,p_i,t_i)
\end{equation}
where $x_i$ and $y_i$ are the pixel positions, $p_i$ is the polarity (the direction of brightness change in the pixel), and $t_i$ is the timestamp. 

In the original version of HOTS, every incoming event gives rise to an instantaneous rise exponential decay kernel with no memory. In our model, incoming events give rise to double exponential decays with STP. We assign a LiWES memristor to each spatial location and polarity $(x,y,p)$, and using the events at $(x,y,p)$ to generate WRITE signals to that memristor, which generate changes in its conductance $G_{x,y,p}(t)$ following the model described in Section \ref{sec:mathmodel}. For each event $ev_i$, we create a time surface by sampling the conductance at all memristors within a square spatial window of lateral size $l$ around $(x_i,y_i)$, i.e. 
\begin{equation} \label{eq:timesurface}
\textit{S}_i(m,n,p) = G_{x_i+m,y_i+n,p}(t_i) 
\end{equation}
for $m, n \in \{-(l-1)/2,\ldots,(l-1)/2\}$ and for all polarities $p$, where the subscripts indicate the memristor's location and polarity.

We use an unsupervised clustering method, such as the K-means algorithm, to cluster the time surfaces. The clusters capture recurring spatio-temporal features of the input data. Each input event generates an output event at the same location and time, but whose polarity is given by the closest cluster. Thus, the number of possible polarities of output events is equal to the number of clusters. 

We can define a multiple layer architecture by defining each layer $k$ as a single iteration of this process. Its input events are
\begin{equation} \label{eq:layer_event}
ev^{k}_{i}= (x^{k}_i,y^{k}_i,p^{k}_i,t^{k}_i)
\end{equation}
Its output events are:
\begin{equation} \label{eq:layer_event+1}
ev^{k+1}_{i}= (x^{k+1}_i,y^{k+1}_i,p^{k+1}_i,t^{k+1}_i)
\end{equation}
where superscripts index the layer number, 
$x^{k+1}_i = x^{k}_i$, $y^{k+1}_i = y^{k}_i$ and $t^{k+1}_i = t^{k}_i$. The output events of one layer become the input to the next. However, to increase spatial integration from layer to layer, we often sub-sample the output events of one layer before inputting them to the next layer. 

Fig. \ref{fig:Architecture}(g)-(m) shows an example of a two-layer architecture. In (g), we plot the most recent input events before a reference event $ev^{1}_i$. We sample the memristor conductances in the square neighborhood around $(x^1_i,y^1_i)$ (shown as the red square) to produce a Time-Surface $S^{1}_i$ (h, only one polarity shown). This Time-Surface gets assigned to a cluster of Layer 1 (i), producing a new output event $ev^{2}_i$ with a new polarity $p^{2}_i$ (j). Due to sub-sampling, time surfaces in layer 2 usually correspond to larger effective neighborhoods in layer 1.
The entire process can be repeated, as shown in Fig. \ref{fig:Architecture}(k,l,m), until we achieve a desired amount of temporal and spatial integration.

Similarly to \cite{lagorce2016hots}, we create a feature vector for each spike activity recording by building a histogram $\mathcal{H}$ of the polarities of spikes from the last layer collected across all pixels and over the entire recording.

We classify the feature vector using a polynomial Support Vector Classifier (SVC). In our comparative experiments seeking to elucidating the effect of different facets of the memristor dynamics on the computed features, we used the simpler Euclidean distance approach proposed in the original HOTS paper \cite{lagorce2016hots}. For each label, we computed a template $\overline{\mathcal{H}}_{label}$ by averaging over all the histograms with that label in the training set. To classify new digits, we performed template matching using the Euclidean distance measure.

\section{Experimental Results}

\subsection{The effects of programmable integration constants on accuracy}

Tables \ref{tab:FixedV} and  \ref{tab:FixedT} show that different pulse settings (Voltage and duration) give rise to different decay time constants. The ability to tune integration constants is fundamental for neuromorphic applications, as spike rates vary between different applications. While these  results do not enable us to model the full relationship between pulse settings and time constants, they do enable us to investigate whether different applications benefit from different time constants. 

We tested the two-layer network shown in Fig. \ref{fig:Architecture}(g to m) on the N-MNIST \cite{NMNIST} and POKERDVS \cite{POKERDVS} datasets. The sub-sampling factor was 7, which reduces the N-MNIST resolution from 28x28 in layer 1 to 4x4 in layer 2 and the POKERDVS resolution from 35x35 in layer 1 to 5x5 in layer 2. The Time-Surface lateral dimensions for N-MNIST results are $l^{[1]}=7$ and, $l^{[2]}=3$ respectively, for the first and second layer. The Time-Surface lateral dimensions for POKERDVS results are $l^{[1]}=5$ and $l^{[2]}=7$. The number of clusters for the N-MNIST results is $N^{[1]}=32$ and $N^{[2]}=96$. The POKERDVS network was smaller with only $N^{[1]}=8$ and $N^{[2]}=64$ clusters.
To eliminate the effect of device stochasticity and enable comparison with other work on these datasets, which typically do not include device stochasticity, we used the ideal model described in Section \ref{sec:mathmodel}. 

Since our implementation of HOTS uses K-means for learning the time surfaces, which requires relatively little data to train, we only use 10\% of the training set for the N-MNIST results. Files were randomly selected at each run. However, our testing results are reported based on performance on the \emph{entire} test set.  

Table \ref{tab:Pulsecomparison} compares the test-set classification accuracies on the two datasets for all the pulse settings listed in Tables \ref{tab:FixedV} and  \ref{tab:FixedT}. These results were calculated over 5 runs on N-MNIST and 10 runs on POKERDVS. We classified with a polynomial support vector machine of order 3. We
report results with additional classifiers in our supplementary
materials. 

Our results show that we obtained the best performance on the N-MNIST and POKERDVS datasets using different pulse parameters, which resulted in very different time constants. The best performance for N-MNIST were obtained for 1V 1ms-long pulses, which gave time constants $\tau_1 = 10$ ms and $\tau_2 = 390ms$. The best performance for POKERDVS were obtained for 200$\mu$s-long pulses with amplitude either 1V ($\tau_1=5$ms, $\tau_2=92ms$) or 4V ($\tau_1=11$ms, $\tau_2=501ms$). These differences highlight the importance of the ability to tune time constants. 

Our LiWES memristive HOTS network achieves state-of-the-art performance on N-MIST (91.27\%), exceeding that reported by Sironi et al. \cite{sironi2018hats} and Iyer et al. \cite{10.3389/fnins.2021.608567}, despite the use of only 10\% of the training data.

\begin{table*}[ht]

    \caption{{percent classification accuracy for different pulse parameters}}
    \centering
     \label{tab:Pulsecomparison}
         \begin{tabular}{l|cccccccc}
          Pulse settings       & (200$\mu$s,1V)     & (200$\mu$s,2V)     & (200$\mu$s,3V)               & (200ms,4V)         & (500$\mu$s,1V)              & (750$\mu$s,1V)      & (1ms,1V)   \\ \hline
          N-MNIST     & $90.90\pm0.22$ & $90.92\pm0.24$ & $91.15\pm0.37$           & $90.93\pm0.21$ & $90.09\pm0.16$          & $90.95\pm0.19$  & $\mathbf{91.27}\pm0.29$ \\
          POKERDVS    & $\mathbf{98.0}\pm3.31$   & $96.50\pm3.90$  & $97.50\pm4.03$           & $\mathbf{98.00}\pm3.32$  & $97.00\pm3.32$   & $96.50\pm3.20$   & $97.50\pm4.03$ \\
         \end{tabular}         
\end{table*}

\subsection{The effects of stochastic dynamics on accuracy}

\begin{figure*}[t]
    \centering
    \includegraphics[scale=0.32]{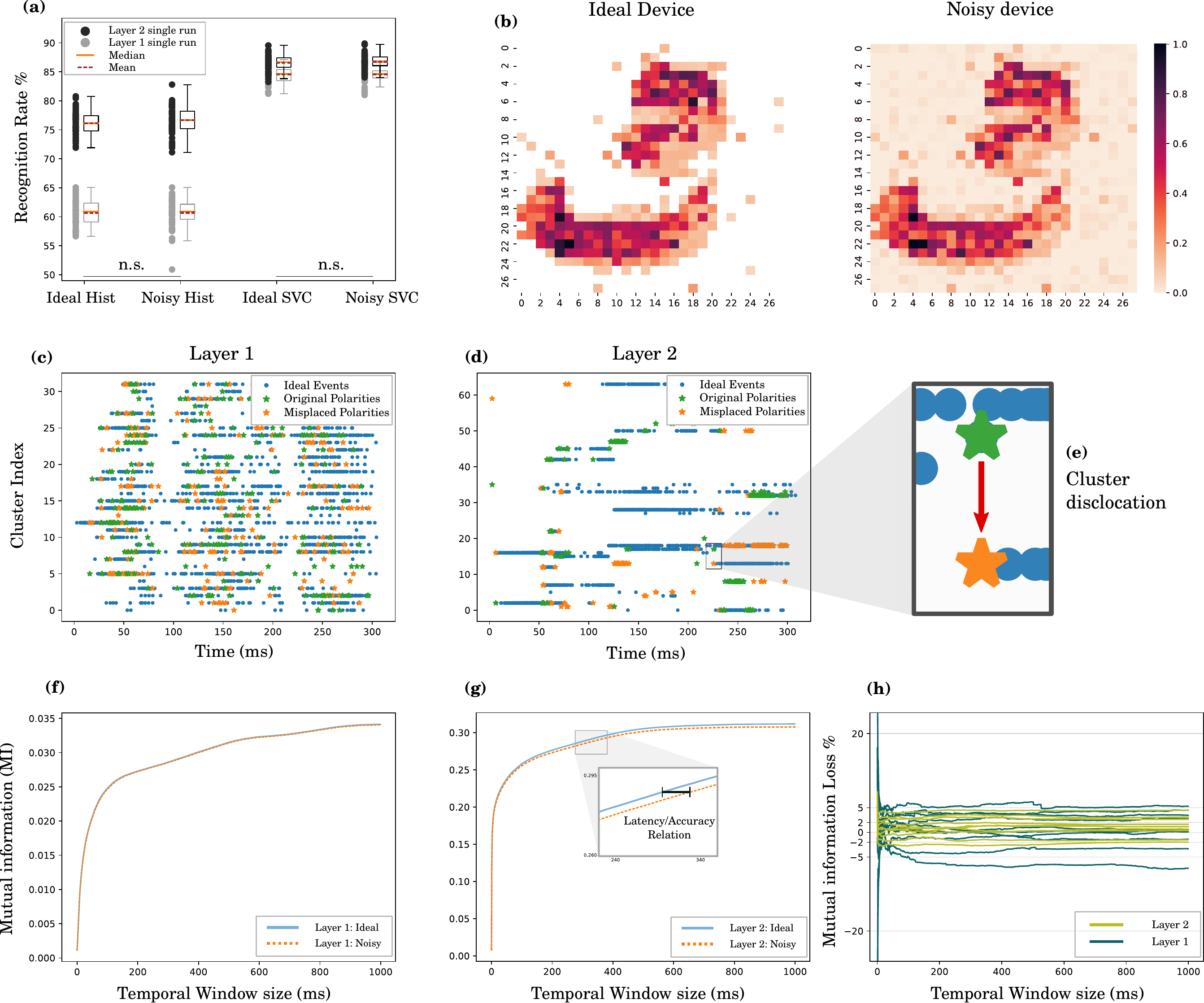}
    \caption{(a) Recognition rates of the Ideal and Noisy networks with the Support Vector and Euclidean Classifiers. (b) A qualitative comparison between Time-Surfaces computed over the entire input using the Ideal and Noisy memristor models. (c and d) The effect of Time-Surface perturbation on cluster (i.e., polarity) assignment in Layers 1 and 2. Blue dots indicate events where the Ideal and Noisy networks make the same cluster assignment. Otherwise, the Ideal (orange dots) and Noisy (green dots) assign events to different clusters. (e) We call this effect cluster dislocation. (f and g) The Mutual Information between the cluster response and the N-MNIST digit labels for Layers 1 and 2 at different Temporal Integration scales. (h) The Mutual Information Percentage Loss due to cluster dislocation. The effect of dislocation is small and constant across all timescales, except for a singularity when the temporal window is zero due to division by zero MI when computing the percentage. More importantly, cluster dislocation is equally likely to increase or decrease Mutual Information.} 
    \label{fig:Noise}
\end{figure*} 

To analyze the effects of stochastic dynamics on recognition rate on a neuromorphic architecture, we compared the performance of HOTS architectures on the N-MNIST dataset using the stochastic memristor model (the 'Noisy' network) and the ideal memristor model (the 'Ideal' network) defined in Section \ref{sec:mathmodel}.

Since the implementation of in-situ learning on the memristive chip was beyond the scope of this paper, we limited noise analysis to inference only. Calculation and clustering of time surfaces was performed using the Ideal network model only. The Noisy network and the Ideal network share the same sets of clusters, but in the case of the Noisy network, the Time-Surfaces generated by the test set were perturbed by the device stochasticity.

The computation time for this test was heavily dependent on the number of clusters and the number of files. For this reason, we set the number of clusters to $N^{[1]}=32$ for layer 1 and $N^{[2]}=64$ for layer 2. We tested only on a random selection of 10\% of the test set every run, and averaged performance over 60 runs.

To ensure the generality of our results, we included results using both the Support Vector Classifier (SVC) and the Euclidean distance classifier (Eucl.). Thus, we considered four cases: Ideal SVC, Noisy SVC, Ideal Eucl. and Noisy Eucl. Each network was trained on the same training sets of N-MNIST data and tested on the same randomly chosen N-MNIST test sets. We also measured classification performance for both layers of the architecture. 

The results in Table \ref{tab:Noise} and Figure \ref{fig:Noise}(a)) show that while different classifiers result in different absolute accuracy, stochasticity in the memristor dynamics does not significantly affect the classification accuracy (t-test $p>0.05$).

However, device stochasticity does influence both Time Surfaces and cluster assignment. 

Fig. \ref{fig:Noise}(b) compares Time Surfaces computed over the entire array ($l=28$) using the Ideal and Noisy models. Although the digit is still recognizable in the Noisy Time Surface, we can clearly see additional variation in the individual pixels. 

As shown in Figs. \ref{fig:Noise}(c) and (d), this variation leads to incorrect cluster (i.e., polarity) assignment of Time-Surfaces, a phenomenon we refer to as 'dislocation' (Fig. \ref{fig:Noise}(e). The mean percentage of events suffering from cluster dislocation in a single run was 10.58\% in Layer 1 and 7.09\% in Layer 2, suggesting that multiple layers might decrease the noise effect. 

One possible explanation for the maintenance of high classification accuracy despite cluster dislocation is that by summarizing events across the entire recording, histogram-based classifiers are "averaging out" the effect of a relatively small number of cluster dislocations. If this were true, then we might expect classifiers integrating information over shorter time scales to be far more affected by dislocation error.

\begin{table}[t]
\caption{Classification accuracy of the Ideal and Noisy Networks with Euclidean and SVC classifiers}
\centering
\begin{tabular}{l|ll}

            & One HOTS Layer       & Two HOTS Layers            \\ \hline
Ideal Eucl & $60.67\%\pm2.21\%$ & $76.12\%\pm1.98\%$   \\ 
Noisy Eucl & $60.69\%\pm2.33\%$ & $76.66\%\pm2.26\%$   \\
Ideal SVC  & $84.54\%\pm1.37\%$ & $86.56\%\pm1.33\%$   \\
Noisy SVC  & $84.55\%\pm1.33\%$ & $86.76\%\pm1.26\%$ 
\end{tabular}
\label{tab:Noise}
\end{table}

To determine whether this is not the case, we calculated the Mutual Information (MI) between events at different time scales and labels, using a method originally presented by Akolkar et al.\cite{akolkar2015can}. In this method, we choose a layer $k$, and sample a random event at that layer from our dataset $ev^l_i$. We create a temporal window of length $\delta$ centered on its timestamp $t^k_i$. We then look across different recordings to calculate the probability of finding another event with the same polarity $p^k_i$ in the same temporal window. We can then calculate the MI between the probability of a response $R$ (equal to 1 if an event with polarity $p^k_i$ is present and 0 otherwise),  $P(R)$, and the probability of the stimulus $S$ (the label of the input), $P(S)$. We repeat this process multiple times including all polarities $p_i$ averaging the result. This value tells us how well the spiking activity  of the layer $k$ at timescale $\delta$ encodes the dataset labels. We include additional information on this method in our Supplementary materials.

By computing the MI at different timescales $\delta$, we can see how well spiking activity at different time scales encodes information about the stimulus labels. Since there is only one label for each recording in the N-MNIST and POKERDVS datasets, we expect the MI to decrease monotonically as the timescale $\delta$ decreases, since shorter timescales contain less information (fewer spikes). 

By comparing the MI computed for the Ideal and Noisy networks, we can see how cluster dislocation affects the MI. If it is true that classifiers integrating information over shorter timescales are more affected by dislocation error, then we would expect the Noisy network's MI to decrease faster than the Ideal network's MI as $delta$ decreases.

Figs. \ref{fig:Noise}(f-g) plot the MI for the Ideal and Noisy networks computed over ten runs using the N-MNIST dataset. As expected, the MI for both networks decreases as $\delta$ decreases. However, the two curves do not diverge as $\delta$ gets smaller, showing the introduction of cluster dislocation due to noise has little effect on the mutual information. We can show this more clearly  using the Mutual Information Percentage Loss, defined by $(MI_{Ideal}-MI_{Noisy})/MI_{Ideal}$. Fig. \ref{fig:Noise}(h) plots the Percentage Loss for each individual run. It remains largely constant across timescale, rarely exceeding 5\% for the first layer and 2\% for the second layer. In addition, the Mutual Information Percentage Loss is equally likely to be positive or negative \ref{fig:Noise}(h). This suggests that rather than causing events to be mapped to less-informative clusters, cluster dislocation often results in events being mapped to equally, if not more, informative clusters.

The MI information loss might be considerably higher for different memristors or datasets. This could affect the accuracy of neuromorphic implementations. One possible solution suggested by this analysis is to exploit the monotonically increasing relationship between timescale and MI shown in Figs. \ref{fig:Noise}(f) and (g), which indicates that the lost information might be recovered by increasing the time window size. However, the window size will negatively impact latency.

\subsection{Computational benefits of Memristive dynamics}

In this section, we investigate whether the more complex dynamics of volatile LiWES memristors bring computational benefits compared to the simpler dynamics assumed in standard HOTS implementations.

We compare the classification accuracy of single-layer HOTS networks built with Ideal Memristor, a simulated Memristor without STP, and two traditional single-decay HOTS architectures. In order to simulate a memristor without STP, we use the ideal model with Eq. (\ref{eq:Lk(t)-noSTP},\ref{eq:Ek(t)-noSTP}), which causes the memristor response to reset to $\overline{A_1} + \overline{A_2}$ at each new incoming event $ev_i$ after the end of the write pulse of width $w$ has been reached. Additionally, we set a single exponential decay model by setting $k=1$, obtaining the original single decay response without STP used for HOTS \cite{lagorce2016hots}. Each network was tested with 30 runs of the N-MNIST dataset. Both the training set and test set were independently sampled for each run. 

\begin{figure}[ht]
\centering
  \includegraphics[width=\columnwidth]{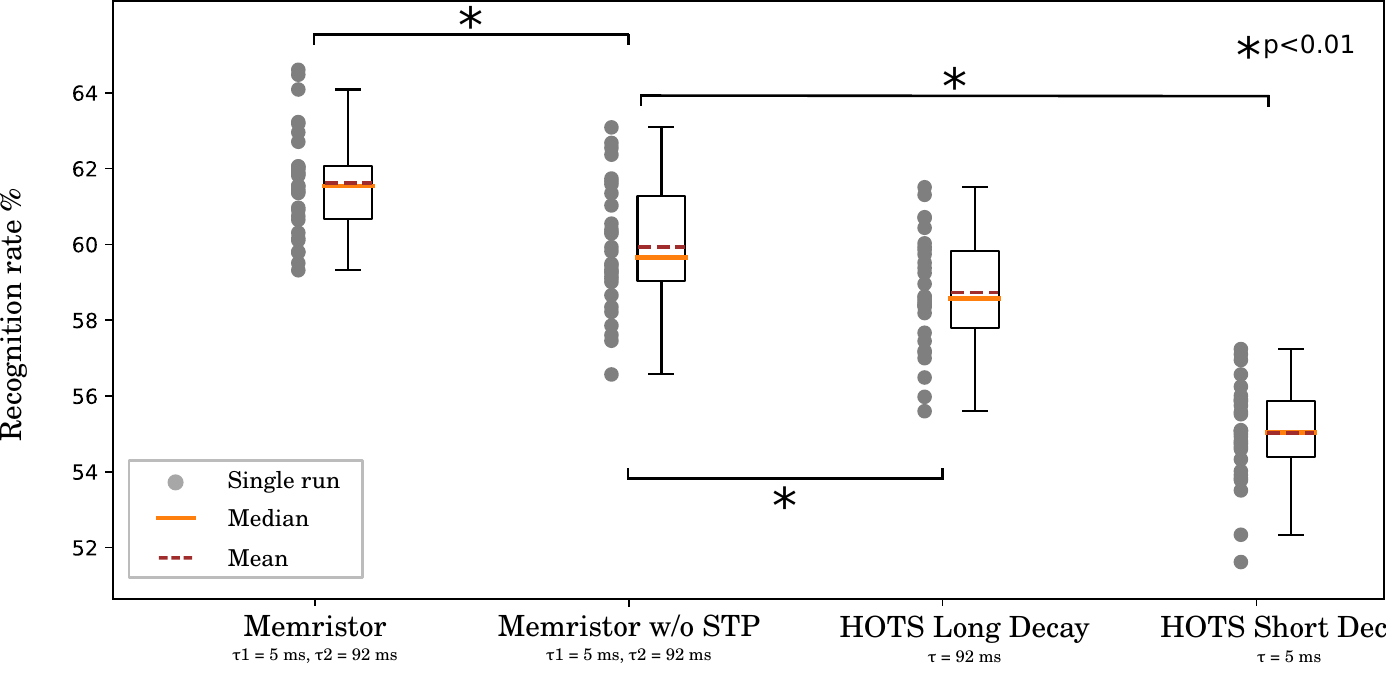}
  \caption{In this test, we compare the recognition rate of our memristor model (with STP and double exponential decay) against the same model without short-term plasticity (STP), HOTS with a single ''long'' decay, or a single ''short'' decay. Both double exponential decay and STP significantly ($p<0.01$) improve accuracy over traditional single-decay w/o STP HOTS.}
  \label{fig:Dynamic_res}
\end{figure}

(Fig \ref{fig:Dynamic_res}) shows the results. For brevity, we only show results with the Euclidean classifier.  Suppl. Table 3 contains additional results. The Memristor model is significantly more accurate than the Memristor without STP and the two single decay HOTS models. Enabling STP results in the largest increase in accuracy. 

\begin{figure*}[ht]
\centering
  \includegraphics[scale=0.26]{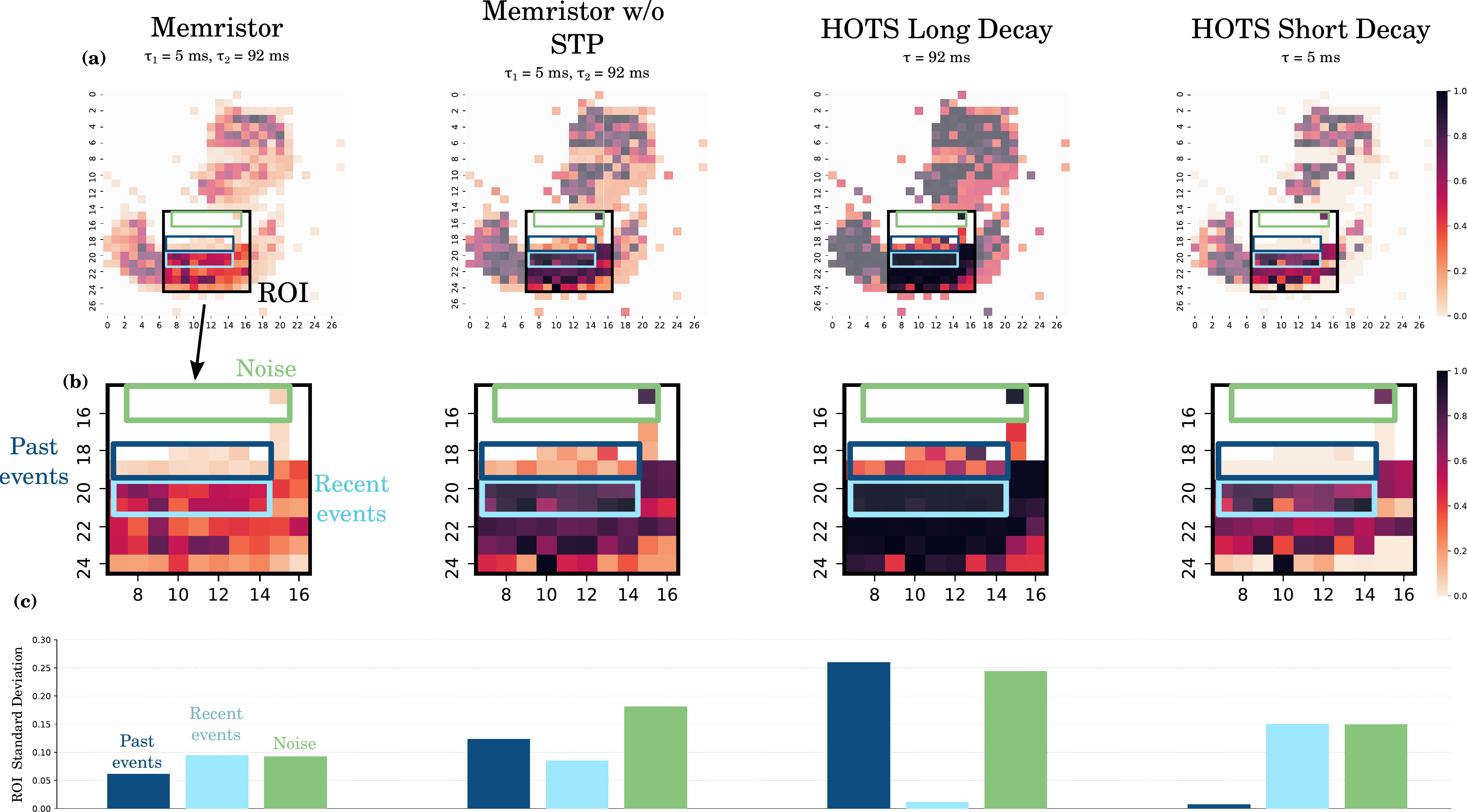}
  \caption{The effect of multiple exponential decays and STP on the Time-Surface representation. (a) Time surfaces for the same input computed using (left to right) HOTS with Ideal Memristor model, HOTS with the Ideal Memristor model but without STP, and the original HOTS with time constants of 5ms and 92ms. (b) Zoom in views of the 11x11 Region Of Interest (ROI), where we identify three sub-regions: recent events (light blue), previous events (dark blue), and sensor noise (green). (c) Plots of the standard deviation of activity in the three regions, which we use as a measure of information and noise. Because of STP, the Memristor model scores the lowest Standard Deviation in the sensor noise region (green). Networks with the double exponential decay (both for the Memristor model and the Memristor w/o STP) have similar standard deviation in windows corresponding to Past and Recent events. In contrast, HOTS single decay models can only represent information at a single timescale, resulting in near zero standard deviations for past events with short decays and for recent  events for long decays.}
  \label{fig:Dynamic_response}
\end{figure*}

Our results also suggest that the double exponential decay better integrates temporal information. Figure \ref{fig:Dynamic_response} shows the effects of STP and double-exponential decays on the time-surface representations. Fig. \ref{fig:Dynamic_response}(a) shows a full digit time-surface at a given time $t_0$ and a 11x11 Region-Of-Interest (ROI) with three distinct sub-regions. The ROIs are plotted in \ref{fig:Dynamic_response}(b), with the sub-regions showing the model response to 'Recent events', 'Past events' and 'Sensor noise'. The last region represents a portion of the frame where the digit is not present. Activity is only caused by the typical salt and pepper noise of the DVS \cite{lichtsteiner2006128}.

Fig. \ref{fig:Dynamic_response}(c) shows the standard deviation of activity in the sub-regions, which is an indirect measure of the amount of information about recent, past, or noise events represented by the time surface. Exponential decay kernels de-emphasize activity that is too fast or slow compared to their decay time-constants. This is evident when compare the standard deviation for recent events (light blue) in the HOTS Long Decay and for past events (dark blue) in the HOTS Short Decay. 

In the original HOTS model \cite{lagorce2016hots}, time surfaces were computed using single exponential decay kernels. Thus, each layer is sensitive to activity only at a single temporal scale. Integration across multiple scales was obtained using multiple layers with increasingly longer time constants. However, the LiWES memristor has double exponential decay response with both short and long time constants. This enables a single layer to integrate information across multiple time scales simultaneously. Thus, the standard deviations of activity for recent and past events are comparable. This is true for both the Memristor model and the Memristor without STP.

The standard deviation from the sensor noise region (in green) achieves its maximum for the HOTS Long Decay model and its minimum for the Memristor model. Longer delays cause the time surfaces to accumulate multiple random events, increasing the standard deviation. In contrast, STP reduces standard deviation by summing the effect of multiple spikes, suppressing the effect of random events.
This is consistent with our finding that the Memristor w/STP has smaller standard deviation in the sensor noise region  compared to the Memristor w/o STP. This effect might also account for our finding that performance of the network is insensitive to device stochasticity. 

\section{Discussion}

Currently, available neuromorphic processors are still in their infancy, as they aim to replicate biological neurons using silicon \cite{7409623, Ning2015, 6809149, 8259423}. However, their application has been limited due to several factors. First, our understanding of the brain is incomplete, lacking a comprehensive theory explaining its operations. Second, the different  physical substrates of silicon and biological brains make it difficult to replicate the fundamental operation of temporal integration in the brain using neuromorphic architectures. Current solutions implement temporal integration digitally \cite{davies2018loihi, painkras2013spinnaker} or through a combination of capacitors and transistors \cite{moradi2017scalable}. In contrast, this work utilizes an electrochemical memristor \cite{Wan_et_moi} with transitory conductance response to implement temporal integration on a single component, opening a path towards the development of compact and energy-efficient neuromorphic systems.

Advancements in technology offer a broader range of materials that could potentially facilitate the design of improved silicon-based brains. Architectures using this device challenge the conventional choices for abstraction level and partitioning in mixed-signal neuromorphic processors. These designs can employ more advanced computation building blocks and design rules. Determining the appropriate level of abstraction \cite{Herz80} remains an open question. The level of abstraction closely interacts with the physical substrate and the computational model design. 

The commonly used level of abstraction, which closely resembles direct biological replication, models neural computation using coupled ordinary differential equations. The temporal dynamics of this model enable information integration over time, while the coupling across state variables models spatial information integration \cite{Abbott1999LapicquesIO}. Analog continuous time VLSI circuits, such as those described by Mead \cite{mead}, are commonly used to implement this level of abstraction. Although these circuits offer low power consumption, they suffer from drawbacks such as mismatch and limited programmability.

A related level of abstraction involves coupled difference equations and is commonly implemented using digital design methodologies in standard CMOS processes or FPGAs \cite{7409623, Ning2015, 6809149, 8259423}. However, even with fully custom designs, these implementations fail to achieve the low power levels sought by neuromorphic engineers. 

Hybrid substrates present an intriguing design space that can leverage the advantages of both analog and digital domains \cite{HybridCMOS}. The most prevalent hybrid model utilizes analog circuits for computation and digital circuits for communication \cite{4633828, CHIU2019255, IndiveriAnalogVLSI}. This approach recognizes that digital circuits operate much faster than the typical spike rate of neurons, enabling a single digital bus to carry signals from multiple neurons. Multi-chip Address Event Representation (AER) networks \cite{7526444, 4541501} embody this level of abstraction, where computation within each chip utilizes analog continuous time circuits, while communication between circuits is digital and often asynchronous.

In this context, we advocate for a hybrid model that employs a different partitioning in the abstraction. Rather than dividing along the lines of function (computation vs. communication), we propose a partitioning based on dimension (time vs. space). We argue that analog implementation is optimal for temporal integration of signals, particularly spiking signals, while digital technologies are better suited for spatial integration. Newly developed memristive technologies \cite{ohno2011stpCBRAM, yang2013filRRAM, sharbati2018Elect, sun20182D} provide an excellent physical substrate for temporal integration. In contrast, spatial integration, which requires signal communication across space, is best achieved using digital technologies.

The memristive network we study moves in the direction of the proposed partitioning. It exhibits space-time separability, as it integrates information over time, pixel by pixel, or more generally, neuron by neuron, followed by spatial integration across pixels or neurons. Space-time separability is a well-known principle in digital signal processing algorithms, offering significant implementation advantages \cite{9156388}. 

However, not all neuromorphic algorithms exhibit space-time separability. For instance, not every set of coupled differential equations can be expressed as a space-time separable set of operations, with the majority being unable to do so.

Nevertheless, we argue that a large and compelling class of algorithms, operating at an abstract scale without relying on spiking neurons, specifically those utilizing time surfaces and hierarchies of time surfaces (HOTS) \cite{lagorce2016hots, sironi2018hats, Grimaldi2021}, naturally exhibit space-time separability as described earlier. These algorithms are ideally suited for implementation using a combination of memristive devices for temporal integration and digital spatial integration, particularly clustering and mapping to the nearest cluster centers. By combining HOTS with a novel three-terminal memristor (Li$_\textbf{x}$WO$_\textbf{3}$), we illustrate how such architectures can be employed for pattern recognition while remaining robust against non-idealities encountered in memristive devices, such as random mismatch and noise.

\section{Conclusion}
Recent developments in semiconductor technology have led to the design and creation of a new class of devices called memristors. It has been predicted that memristors will be used in the near future as the atomic component of more advanced and complex systems, which can provide performance superior to conventional transistor-based hardware \cite{williams08}. In neuromorphic engineering, memristors are more commonly used as "static"  synaptic weights for the spatial integration of signals. While the temporal dynamics of memristors are known in the literature, we still need to understand their computational properties better to be able to exploit them fully in practical scenarios. 

In this work, we used a Li$_\textbf{x}$WO$_\textbf{3}$ electrochemical memristor to test the effect of programmable time constants, double exponential decay and STP on the widely used N-MNIST dataset and POKERDVS dataset. We used single-pulse recordings to build a model of the device and then used it to simulate a HOTS network. The ability to program time constants is important, as different tasks generate spike activity with different temporal dynamics as evidenced by our results comparing the best pulse parameters for the M-NIST and POKERDVS datasets. In this work, we assumed that time constants for all memristors in the network had the same statistics, but moving forward, it may be interesting to investigate setting time constants layer by layer or even neuron by neuron. We showed that the intrinsic stochasticity of the device did not impact accuracy (VI.B  section). However, we also showed a relationship between latency and accuracy that could be used to offset accuracy loss by increasing integration time in devices with less precise temporal dynamics. This is especially important as it extends our considerations to memristors other than the one we tested, such as two-terminal memristors that also exhibit STP \cite{ohno2011stpCBRAM, sun20182D}.

One limitation of our results is that we could not include device mismatch in our simulation, as we have yet to realize a LiWES array, which would allow us to characterize this mismatch. However, we have modelled cycle-to-cycle variation, which also gives rise to mismatch in time-surfaces, albeit over time rather than space. Nonetheless, our results showing robustness to this type of mismatch suggest that our network may also be robust to spatial mismatch. This is a promising avenue for future work.

The last section (VI.C) explored the computational properties of STP and the double exponential dynamics. Both STP and double exponential decay dynamics increased the accuracy of the network compared to the original HOTS network with single exponentials and no-STP. STP contributes to reducing "Noise" in the network. Multiple exponential dynamics allow temporal integration across a broader time scales. Since time surfaces are based on an exponential decay kernel, akin to biological EPSP/IPSPs, we expect our results to generalize to a wider class of models, such as integrate and fire neurons. These results are of particular importance as they highlight the practical use of less explored properties of this class of memristive devices and allow us to envision a future where memristors are used for temporal data processing and synaptic weight, eliminating the need for more complex analog or digital circuits.

Taken as a whole, our work provides strong evidence that the volatile properties of memristors can become a powerful tool for building a more specialized class of neuromorphic systems while reducing design complexity.

\section*{Acknowledgments}
We would like to thank Dr. Sio-Hoi Ieng for his valuable comments and discussion during the manuscript preparation, and Richard Newcombe for supporting this work.


\bibliographystyle{IEEEtran}

\bibliography{Biblio.bib}  

\begin{thebibliography}{10}
\providecommand{\url}[1]{#1}
\csname url@samestyle\endcsname
\providecommand{\newblock}{\relax}
\providecommand{\bibinfo}[2]{#2}
\providecommand{\BIBentrySTDinterwordspacing}{\spaceskip=0pt\relax}
\providecommand{\BIBentryALTinterwordstretchfactor}{4}
\providecommand{\BIBentryALTinterwordspacing}{\spaceskip=\fontdimen2\font plus
\BIBentryALTinterwordstretchfactor\fontdimen3\font minus \fontdimen4\font\relax}
\providecommand{\BIBforeignlanguage}[2]{{%
\expandafter\ifx\csname l@#1\endcsname\relax
\typeout{** WARNING: IEEEtran.bst: No hyphenation pattern has been}%
\typeout{** loaded for the language `#1'. Using the pattern for}%
\typeout{** the default language instead.}%
\else
\language=\csname l@#1\endcsname
\fi
#2}}
\providecommand{\BIBdecl}{\relax}
\BIBdecl

\bibitem{Pierson2017}
\BIBentryALTinterwordspacing
H.~A. Pierson and M.~S. Gashler, ``Deep learning in robotics: a review of recent research,'' \emph{Advanced Robotics}, vol.~31, no.~16, pp. 821--835, 2017. [Online]. Available: \url{https://doi.org/10.1080/01691864.2017.1365009}
\BIBentrySTDinterwordspacing

\bibitem{datascaling_chinchilla}
J.~Hoffmann, S.~Borgeaud, A.~Mensch, E.~Buchatskaya, T.~Cai, E.~Rutherford, D.~d.~L. Casas, L.~A. Hendricks, J.~Welbl, A.~Clark \emph{et~al.}, ``Training compute-optimal large language models,'' \emph{arXiv preprint arXiv:2203.15556}, 2022.

\bibitem{schuman2017survey}
C.~D. Schuman, T.~E. Potok, R.~M. Patton, J.~D. Birdwell, M.~E. Dean, G.~S. Rose, and J.~S. Plank, ``A survey of neuromorphic computing and neural networks in hardware,'' \emph{arXiv preprint arXiv:1705.06963}, 2017.

\bibitem{lichtsteiner2006128}
P.~Lichtsteiner, C.~Posch, and T.~Delbruck, ``A 128 x 128 120db 30mw asynchronous vision sensor that responds to relative intensity change,'' in \emph{2006 IEEE International Solid State Circuits Conference-Digest of Technical Papers}.\hskip 1em plus 0.5em minus 0.4em\relax IEEE, 2006, pp. 2060--2069.

\bibitem{2011SequenceMemristor}
D.~{Kuzum}, R.~G.~D. {Jeyasingh}, and H.~.~P. {Wong}, ``Energy efficient programming of nanoelectronic synaptic devices for large-scale implementation of associative and temporal sequence learning,'' in \emph{2011 International Electron Devices Meeting}, 2011, pp. 30.3.1--30.3.4.

\bibitem{2011PCMComplexVisualPatterns}
M.~Suri, O.~Bichler, D.~Querlioz, O.~Cueto, L.~Perniola, V.~Sousa, D.~Vuillaume, C.~Gamrat, and B.~DeSalvo, ``Phase change memory as synapse for ultra-dense neuromorphic systems: Application to complex visual pattern extraction,'' in \emph{2011 International Electron Devices Meeting}.\hskip 1em plus 0.5em minus 0.4em\relax IEEE, 2011, pp. 4--4.

\bibitem{Ambrogio_2013_RRAM}
\BIBentryALTinterwordspacing
S.~Ambrogio, S.~Balatti, F.~Nardi, S.~Facchinetti, and D.~Ielmini, ``Spike-timing dependent plasticity in a transistor-selected resistive switching memory,'' \emph{Nanotechnology}, vol.~24, no.~38, p. 384012, 9 2013. [Online]. Available: \url{https://doi.org/10.1088%2F0957-4484%2F24%2F38%2F384012}
\BIBentrySTDinterwordspacing

\bibitem{Recursive2014PCM}
\BIBentryALTinterwordspacing
S.~B. Eryilmaz, D.~Kuzum, R.~Jeyasingh, S.~Kim, M.~BrightSky, C.~Lam, and H.-S.~P. Wong, ``Brain-like associative learning using a nanoscale non-volatile phase change synaptic device array,'' \emph{Frontiers in Neuroscience}, vol.~8, p. 205, 2014. [Online]. Available: \url{https://www.frontiersin.org/article/10.3389/fnins.2014.00205}
\BIBentrySTDinterwordspacing

\bibitem{2015STT-MRAMNetwork}
A.~F. {Vincent}, J.~{Larroque}, N.~{Locatelli}, N.~{Ben Romdhane}, O.~{Bichler}, C.~{Gamrat}, W.~S. {Zhao}, J.~{Klein}, S.~{Galdin-Retailleau}, and D.~{Querlioz}, ``Spin-transfer torque magnetic memory as a stochastic memristive synapse for neuromorphic systems,'' \emph{IEEE Transactions on Biomedical Circuits and Systems}, vol.~9, no.~2, pp. 166--174, 2015.

\bibitem{Kmeans2018}
\BIBentryALTinterwordspacing
Y.~Jeong, J.~Lee, J.~Moon, J.~H. Shin, and W.~D. Lu, ``K-means data clustering with memristor networks,'' \emph{Nano Letters}, vol.~18, no.~7, pp. 4447--4453, 2018, pMID: 29879355. [Online]. Available: \url{https://doi.org/10.1021/acs.nanolett.8b01526}
\BIBentrySTDinterwordspacing

\bibitem{wang2018fully}
Z.~Wang, S.~Joshi, S.~Savel’ev, W.~Song, R.~Midya, Y.~Li, M.~Rao, P.~Yan, S.~Asapu, Y.~Zhuo \emph{et~al.}, ``Fully memristive neural networks for pattern classification with unsupervised learning,'' \emph{Nature Electronics}, vol.~1, no.~2, pp. 137--145, 2018.

\bibitem{Xia2019}
Q.~Xia and J.~J. Yang, ``Memristive crossbar arrays for brain-inspired computing,'' \emph{Nature Materials}, vol.~18, pp. 309--323, 4 2019.

\bibitem{wan2019emerging}
Q.~Wan, M.~T. Sharbati, J.~R. Erickson, Y.~Du, and F.~Xiong, ``Emerging artificial synaptic devices for neuromorphic computing,'' \emph{Advanced Materials Technologies}, vol.~4, p. 1900037, 4 2019.

\bibitem{yang2013filRRAM}
R.~Yang, K.~Terabe, Y.~Yao, T.~Tsuruoka, T.~Hasegawa, J.~K. Gimzewski, and M.~Aono, ``Synaptic plasticity and memory functions achieved in a wo3- x-based nanoionics device by using the principle of atomic switch operation,'' \emph{Nanotechnology}, vol.~24, p. 384003, 2013.

\bibitem{ohno2011stpCBRAM}
T.~Ohno, T.~Hasegawa, T.~Tsuruoka, K.~Terabe, J.~K. Gimzewski, and M.~Aono, ``Short-term plasticity and long-term potentiation mimicked in single inorganic synapses,'' \emph{Nature materials}, vol.~10, pp. 591--595, 2011.

\bibitem{berdan2016emulating}
R.~Berdan, E.~Vasilaki, A.~Khiat, G.~Indiveri, A.~Serb, and T.~Prodromakis, ``Emulating short-term synaptic dynamics with memristive devices,'' \emph{Scientific reports}, vol.~6, pp. 1--9, 2016.

\bibitem{sharbati2018Elect}
M.~T. Sharbati, Y.~Du, J.~Torres, N.~D. Ardolino, M.~Yun, and F.~Xiong, ``Low-power, electrochemically tunable graphene synapses for neuromorphic computing,'' \emph{Advanced Materials}, vol.~30, no.~36, p. 1802353, 2018.

\bibitem{double_decay_mem}
J.~Zhu, Y.~Yang, R.~Jia, Z.~Liang, W.~Zhu, Z.~U. Rehman, L.~Bao, X.~Zhang, Y.~Cai, L.~Song \emph{et~al.}, ``Ion gated synaptic transistors based on 2d van der waals crystals with tunable diffusive dynamics,'' \emph{Advanced Materials}, vol.~30, no.~21, p. 1800195, 2018.

\bibitem{Hua2019}
Q.~Hua, H.~Wu, B.~Gao, Q.~Zhang, W.~Wu, Y.~Li, X.~Wang, W.~Hu, and H.~Qian, ``Low‐voltage oscillatory neurons for memristor‐based neuromorphic systems,'' \emph{Global Challenges}, vol.~3, p. 1900015, 11 2019.

\bibitem{Lim2016}
H.~Lim, H.-W. Ahn, V.~Kornijcuk, G.~Kim, J.~Y. Seok, I.~Kim, C.~S. Hwang, and D.~S. Jeong, ``Relaxation oscillator-realized artificial electronic neurons, their responses, and noise,'' \emph{Nanoscale}, vol.~8, pp. 9629--9640, 2016.

\bibitem{Pickett2013}
M.~D. Pickett, G.~Medeiros-Ribeiro, and R.~S. Williams, ``A scalable neuristor built with mott memristors,'' \emph{Nature Materials}, vol.~12, pp. 114--117, 2 2013.

\bibitem{Sung2022}
S.~H. Sung, T.~J. Kim, H.~Shin, T.~H. Im, and K.~J. Lee, ``Simultaneous emulation of synaptic and intrinsic plasticity using a memristive synapse,'' \emph{Nature Communications}, vol.~13, p. 2811, 12 2022.

\bibitem{Sarwat2022}
S.~G. Sarwat, B.~Kersting, T.~Moraitis, V.~P. Jonnalagadda, and A.~Sebastian, ``Phase-change memtransistive synapses for mixed-plasticity neural computations,'' \emph{Nature Nanotechnology}, vol.~17, pp. 507--513, 5 2022.

\bibitem{Wan_et_moi}
\BIBentryALTinterwordspacing
Q.~Wan, M.~Rasetto, M.~T. Sharbati, J.~R. Erickson, S.~R. Velagala, M.~T. Reilly, Y.~Li, R.~Benosman, and F.~Xiong, ``Low-voltage electrochemical lixwo3 synapses with temporal dynamics for spiking neural networks,'' \emph{Advanced Intelligent Systems}, vol. n/a, p. 2100021, 2021. [Online]. Available: \url{https://onlinelibrary.wiley.com/doi/abs/10.1002/aisy.202100021}
\BIBentrySTDinterwordspacing

\bibitem{sun20182D}
L.~Sun, Y.~Zhang, G.~Hwang, J.~Jiang, D.~Kim, Y.~A. Eshete, R.~Zhao, and H.~Yang, ``Synaptic computation enabled by joule heating of single-layered semiconductors for sound localization,'' \emph{Nano Letters}, vol.~18, pp. 3229--3234, 2018.

\bibitem{shaban2021adaptive}
A.~Shaban, S.~S. Bezugam, and M.~Suri, ``An adaptive threshold neuron for recurrent spiking neural networks with nanodevice hardware implementation,'' \emph{Nature Communications}, vol.~12, no.~1, p. 4234, 2021.

\bibitem{demiraug2021pcm}
Y.~Demira{\u{g}}, F.~Moro, T.~Dalgaty, G.~Navarro, C.~Frenkel, G.~Indiveri, E.~Vianello, and M.~Payvand, ``Pcm-trace: scalable synaptic eligibility traces with resistivity drift of phase-change materials,'' in \emph{2021 IEEE International Symposium on Circuits and Systems (ISCAS)}.\hskip 1em plus 0.5em minus 0.4em\relax IEEE, 2021, pp. 1--5.

\bibitem{lagorce2016hots}
X.~Lagorce, G.~Orchard, F.~Galluppi, B.~E. Shi, and R.~B. Benosman, ``Hots: a hierarchy of event-based time-surfaces for pattern recognition,'' \emph{IEEE transactions on pattern analysis and machine intelligence}, vol.~39, no.~7, pp. 1346--1359, 2016.

\bibitem{NMNIST}
G.~Orchard, A.~Jayawant, G.~K. Cohen, and N.~Thakor, ``Converting static image datasets to spiking neuromorphic datasets using saccades,'' \emph{Frontiers in neuroscience}, vol.~9, p. 437, 2015.

\bibitem{POKERDVS}
T.~Serrano-Gotarredona and B.~Linares-Barranco, ``Poker-dvs and mnist-dvs. their history, how they were made, and other details,'' \emph{Frontiers in neuroscience}, vol.~9, p. 481, 2015.

\bibitem{FirstLi}
E.~J. Fuller, F.~E. Gabaly, F.~L{\'e}onard, S.~Agarwal, S.~J. Plimpton, R.~B. Jacobs-Gedrim, C.~D. James, M.~J. Marinella, and A.~A. Talin, ``Li-ion synaptic transistor for low power analog computing,'' \emph{Advanced Materials}, vol.~29, no.~4, p. 1604310, 2017.

\bibitem{li2019low}
Y.~Li, E.~J. Fuller, S.~Asapu, S.~Agarwal, T.~Kurita, J.~J. Yang, and A.~A. Talin, ``Low-voltage, cmos-free synaptic memory based on li x tio2 redox transistors,'' \emph{ACS applied materials \& interfaces}, vol.~11, no.~42, pp. 38\,982--38\,992, 2019.

\bibitem{fuller2019parallel}
E.~J. Fuller, S.~T. Keene, A.~Melianas, Z.~Wang, S.~Agarwal, Y.~Li, Y.~Tuchman, C.~D. James, M.~J. Marinella, J.~J. Yang \emph{et~al.}, ``Parallel programming of an ionic floating-gate memory array for scalable neuromorphic computing,'' \emph{Science}, vol. 364, no. 6440, pp. 570--574, 2019.

\bibitem{yang2018all}
C.-S. Yang, D.-S. Shang, N.~Liu, E.~J. Fuller, S.~Agrawal, A.~A. Talin, Y.-Q. Li, B.-G. Shen, and Y.~Sun, ``All-solid-state synaptic transistor with ultralow conductance for neuromorphic computing,'' \emph{Advanced Functional Materials}, vol.~28, no.~42, p. 1804170, 2018.

\bibitem{qian2016artificial}
C.~Qian, J.~Sun, L.-a. Kong, G.~Gou, J.~Yang, J.~He, Y.~Gao, and Q.~Wan, ``Artificial synapses based on in-plane gate organic electrochemical transistors,'' \emph{ACS applied materials \& interfaces}, vol.~8, no.~39, pp. 26\,169--26\,175, 2016.

\bibitem{Wan_dynamics}
\BIBentryALTinterwordspacing
Q.~Wan, P.~Zhang, Q.~Shao, M.~T. Sharbati, J.~R. Erickson, K.~L. Wang, and F.~Xiong, ``(bi0.2sb0.8)2te3 based dynamic synapses with programmable spatio-temporal dynamics,'' \emph{APL Materials}, vol.~7, no.~10, p. 101107, 2019. [Online]. Available: \url{https://doi.org/10.1063/1.5106381}
\BIBentrySTDinterwordspacing

\bibitem{van2017non}
Y.~van~de Burgt, E.~Lubberman, E.~J. Fuller, S.~T. Keene, G.~C. Faria, S.~Agarwal, M.~J. Marinella, A.~A. Talin, and A.~Salleo, ``A non-volatile organic electrochemical device as a low-voltage artificial synapse for neuromorphic computing,'' \emph{Nature materials}, vol.~16, no.~4, pp. 414--418, 2017.

\bibitem{yao2020protonic}
X.~Yao, K.~Klyukin, W.~Lu, M.~Onen, S.~Ryu, D.~Kim, N.~Emond, I.~Waluyo, A.~Hunt, J.~A. Del~Alamo \emph{et~al.}, ``Protonic solid-state electrochemical synapse for physical neural networks,'' \emph{Nature communications}, vol.~11, no.~1, pp. 1--10, 2020.

\bibitem{li2021one}
Y.~Li, Z.~Xuan, J.~Lu, Z.~Wang, X.~Zhang, Z.~Wu, Y.~Wang, H.~Xu, C.~Dou, Y.~Kang \emph{et~al.}, ``One transistor one electrolyte-gated transistor based spiking neural network for power-efficient neuromorphic computing system,'' \emph{Advanced Functional Materials}, p. 2100042, 2021.

\bibitem{li2020filament}
Y.~Li, E.~J. Fuller, J.~D. Sugar, S.~Yoo, D.~S. Ashby, C.~H. Bennett, R.~D. Horton, M.~S. Bartsch, M.~J. Marinella, W.~D. Lu \emph{et~al.}, ``Filament-free bulk resistive memory enables deterministic analogue switching,'' \emph{Advanced Materials}, vol.~32, no.~45, p. 2003984, 2020.

\bibitem{tang2018ecram}
J.~Tang, D.~Bishop, S.~Kim, M.~Copel, T.~Gokmen, T.~Todorov, S.~Shin, K.-T. Lee, P.~Solomon, K.~Chan \emph{et~al.}, ``Ecram as scalable synaptic cell for high-speed, low-power neuromorphic computing,'' in \emph{2018 IEEE International Electron Devices Meeting (IEDM)}.\hskip 1em plus 0.5em minus 0.4em\relax IEEE, 2018, pp. 13--1.

\bibitem{williams1970non}
G.~Williams and D.~C. Watts, ``Non-symmetrical dielectric relaxation behaviour arising from a simple empirical decay function,'' \emph{Transactions of the Faraday society}, vol.~66, pp. 80--85, 1970.

\bibitem{li2017electric}
H.-M. Li, K.~Xu, B.~Bourdon, H.~Lu, Y.-C. Lin, J.~A. Robinson, A.~C. Seabaugh, and S.~K. Fullerton-Shirey, ``Electric double layer dynamics in poly (ethylene oxide) liclo4 on graphene transistors,'' \emph{The Journal of Physical Chemistry C}, vol. 121, no.~31, pp. 16\,996--17\,004, 2017.

\bibitem{purves2001neuroscience}
\BIBentryALTinterwordspacing
D.~Purves, G.~Augustine, D.~Fitzpatrick, L.~Katz, A.~LaMantia, J.~McNamara, and S.~Williams, ``Neuroscience 2nd edition,'' in \emph{Neuroscience 2nd edition}.\hskip 1em plus 0.5em minus 0.4em\relax Sinauer Associates, 2001, ch. 7: Two Families of Postsynaptic Receptors. [Online]. Available: \url{https://www.ncbi.nlm.nih.gov/books/NBK10855/}
\BIBentrySTDinterwordspacing

\bibitem{stereoHOTS}
\BIBentryALTinterwordspacing
S.-H. Ieng, J.~Carneiro, M.~Osswald, and R.~Benosman, ``{Neuromorphic Event-Based Generalized Time-Based Stereovision},'' \emph{Frontiers in Neuroscience}, vol.~12, p. 442, 2018. [Online]. Available: \url{https://www.frontiersin.org/article/10.3389/fnins.2018.00442}
\BIBentrySTDinterwordspacing

\bibitem{gauvain2021optogenetic}
G.~Gauvain, H.~Akolkar, A.~Chaffiol, F.~Arcizet, M.~A. Khoei, M.~Desrosiers, C.~Jaillard, R.~Caplette, O.~Marre, S.~Bertin \emph{et~al.}, ``Optogenetic therapy: high spatiotemporal resolution and pattern discrimination compatible with vision restoration in non-human primates,'' \emph{Communications biology}, vol.~4, no.~1, pp. 1--15, 2021.

\bibitem{rasetto2021event}
M.~Rasetto, J.~P. Dominguez-Morales, A.~Jimenez-Fernandez, and R.~Benosman, ``Event based time-vectors for auditory features extraction: a neuromorphic approach for low power audio recognition,'' \emph{arXiv preprint arXiv:2112.07011}, 2021.

\bibitem{sironi2018hats}
A.~Sironi, M.~Brambilla, N.~Bourdis, X.~Lagorce, and R.~Benosman, ``Hats: Histograms of averaged time surfaces for robust event-based object classification,'' in \emph{Proceedings of the IEEE Conference on Computer Vision and Pattern Recognition}, 2018, pp. 1731--1740.

\bibitem{10.3389/fnins.2021.608567}
\BIBentryALTinterwordspacing
L.~R. Iyer, Y.~Chua, and H.~Li, ``Is neuromorphic mnist neuromorphic? analyzing the discriminative power of neuromorphic datasets in the time domain,'' \emph{Frontiers in Neuroscience}, vol.~15, 2021. [Online]. Available: \url{https://www.frontiersin.org/articles/10.3389/fnins.2021.608567}
\BIBentrySTDinterwordspacing

\bibitem{akolkar2015can}
H.~Akolkar, C.~Meyer, X.~Clady, O.~Marre, C.~Bartolozzi, S.~Panzeri, and R.~Benosman, ``What can neuromorphic event-driven precise timing add to spike-based pattern recognition?'' \emph{Neural computation}, vol.~27, no.~3, pp. 561--593, 2015.

\bibitem{7409623}
G.~Indiveri, F.~Corradi, and N.~Qiao, ``Neuromorphic architectures for spiking deep neural networks,'' in \emph{2015 IEEE International Electron Devices Meeting (IEDM)}, 2015, pp. 4.2.1--4.2.4.

\bibitem{Ning2015}
\BIBentryALTinterwordspacing
N.~Qiao, H.~Mostafa, F.~Corradi, M.~Osswald, F.~Stefanini, D.~Sumislawska, and G.~Indiveri, ``A reconfigurable on-line learning spiking neuromorphic processor comprising 256 neurons and 128k synapses,'' \emph{Frontiers in Neuroscience}, vol.~9, p. 141, 2015. [Online]. Available: \url{https://www.frontiersin.org/article/10.3389/fnins.2015.00141}
\BIBentrySTDinterwordspacing

\bibitem{6809149}
E.~Chicca, F.~Stefanini, C.~Bartolozzi, and G.~Indiveri, ``Neuromorphic electronic circuits for building autonomous cognitive systems,'' \emph{Proceedings of the IEEE}, vol. 102, no.~9, pp. 1367--1388, 2014.

\bibitem{8259423}
M.~Davies, N.~Srinivasa, T.-H. Lin, G.~Chinya, Y.~Cao, S.~H. Choday, G.~Dimou, P.~Joshi, N.~Imam, S.~Jain, Y.~Liao, C.-K. Lin, A.~Lines, R.~Liu, D.~Mathaikutty, S.~McCoy, A.~Paul, J.~Tse, G.~Venkataramanan, Y.-H. Weng, A.~Wild, Y.~Yang, and H.~Wang, ``Loihi: A neuromorphic manycore processor with on-chip learning,'' \emph{IEEE Micro}, vol.~38, no.~1, pp. 82--99, 2018.

\bibitem{davies2018loihi}
M.~Davies, N.~Srinivasa, T.-H. Lin, G.~Chinya, Y.~Cao, S.~H. Choday, G.~Dimou, P.~Joshi, N.~Imam, S.~Jain \emph{et~al.}, ``Loihi: A neuromorphic manycore processor with on-chip learning,'' \emph{Ieee Micro}, vol.~38, no.~1, pp. 82--99, 2018.

\bibitem{painkras2013spinnaker}
E.~Painkras, L.~A. Plana, J.~Garside, S.~Temple, F.~Galluppi, C.~Patterson, D.~R. Lester, A.~D. Brown, and S.~B. Furber, ``Spinnaker: A 1-w 18-core system-on-chip for massively-parallel neural network simulation,'' \emph{IEEE Journal of Solid-State Circuits}, vol.~48, no.~8, pp. 1943--1953, 2013.

\bibitem{moradi2017scalable}
S.~Moradi, N.~Qiao, F.~Stefanini, and G.~Indiveri, ``A scalable multicore architecture with heterogeneous memory structures for dynamic neuromorphic asynchronous processors (dynaps),'' \emph{IEEE transactions on biomedical circuits and systems}, vol.~12, no.~1, pp. 106--122, 2017.

\bibitem{Herz80}
\BIBentryALTinterwordspacing
A.~V.~M. Herz, T.~Gollisch, C.~K. Machens, and D.~Jaeger, ``Modeling single-neuron dynamics and computations: A balance of detail and abstraction,'' \emph{Science}, vol. 314, no. 5796, pp. 80--85, 2006. [Online]. Available: \url{https://science.sciencemag.org/content/314/5796/80}
\BIBentrySTDinterwordspacing

\bibitem{Abbott1999LapicquesIO}
L.~Abbott, ``Lapicque’s introduction of the integrate-and-fire model neuron (1907),'' \emph{Brain Research Bulletin}, vol.~50, pp. 303--304, 1999.

\bibitem{mead}
C.~Mead and I.~Mohammed, \emph{Analog VLSI Implementation of Neural Systems}, 1989, vol.~80.

\bibitem{HybridCMOS}
D.~B. Strukov, D.~R. Stewart, J.~Borghetti, X.~Li, M.~Pickett, G.~M. Ribeiro, W.~Robinett, G.~Snider, J.~P. Strachan, W.~Wu, Q.~Xia, J.~J. Yang, and R.~S. Williams, ``Hybrid cmos/memristor circuits,'' in \emph{2010 IEEE International Symposium on Circuits and Systems (ISCAS)}, 2010, pp. 1967--1970.

\bibitem{4633828}
J.~Schemmel, J.~Fieres, and K.~Meier, ``Wafer-scale integration of analog neural networks,'' in \emph{2008 IEEE International Joint Conference on Neural Networks (IEEE World Congress on Computational Intelligence)}, 2008, pp. 431--438.

\bibitem{CHIU2019255}
\BIBentryALTinterwordspacing
R.~Chiu, D.~López-Mancilla, C.~E. Castañeda, O.~Orozco-López, E.~Villafaña-Rauda, and R.~Sevilla-Escoboza, ``Design and implementation of a jerk circuit using a hybrid analog–digital system,'' \emph{Chaos, Solitons and Fractals}, vol. 119, pp. 255--262, 2019. [Online]. Available: \url{https://www.sciencedirect.com/science/article/pii/S0960077919300062}
\BIBentrySTDinterwordspacing

\bibitem{IndiveriAnalogVLSI}
\BIBentryALTinterwordspacing
G.~Indiveri, ``{Modeling Selective Attention Using a Neuromorphic Analog VLSI Device},'' \emph{Neural Computation}, vol.~12, no.~12, pp. 2857--2880, 12 2000. [Online]. Available: \url{https://doi.org/10.1162/089976600300014755}
\BIBentrySTDinterwordspacing

\bibitem{7526444}
J.~Park, T.~Yu, S.~Joshi, C.~Maier, and G.~Cauwenberghs, ``Hierarchical address event routing for reconfigurable large-scale neuromorphic systems,'' \emph{IEEE Transactions on Neural Networks and Learning Systems}, vol.~28, no.~10, pp. 2408--2422, 2017.

\bibitem{4541501}
D.~B. Fasnacht, A.~M. Whatley, and G.~Indiveri, ``A serial communication infrastructure for multi-chip address event systems,'' in \emph{2008 IEEE International Symposium on Circuits and Systems}, 2008, pp. 648--651.

\bibitem{9156388}
M.~Haris, G.~Shakhnarovich, and N.~Ukita, ``Space-time-aware multi-resolution video enhancement,'' in \emph{2020 IEEE/CVF Conference on Computer Vision and Pattern Recognition (CVPR)}, 2020, pp. 2856--2865.

\bibitem{Grimaldi2021}
A.~Grimaldi, V.~Boutin, L.~Perrinet, S.-H. Ieng, and R.~Benosman, ``A homeostatic gain control mechanism to improve event-driven object recognition,'' in \emph{2021 International Conference on Content-Based Multimedia Indexing (CBMI)}, 2021, pp. 1--6.

\bibitem{williams08}
D.~Strukov and S.~Williams, ``Exponential ionic drift: Fast switching and low volatility of thin-film memristors,'' \emph{Applied Physics A}, vol.~94, pp. 515--519, 03 2009.

\end{thebibliography}
%

\begin{IEEEbiography}[{\includegraphics[width=1in,height=1.25in,clip,keepaspectratio]{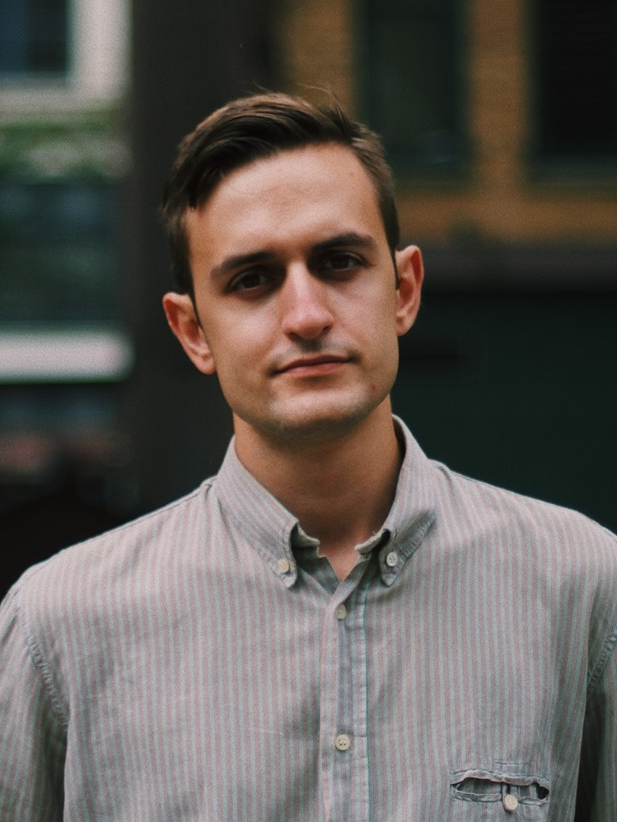}}]{Marco Rasetto}
received the M.Sc and B.Sc degrees in Bioengineering at the University of Genova, Italy, in 2016 and 2018, respectively. His research interests include neuromorphic algorithms, computational neuroscience, machine learning, and computer vision, with a focus on efficient sensory systems for edge computing. He is currently a Ph.D. candidate in Bioengineering at the University of Pittsburgh and the Center for Neural Basis of Cognition, Pennsylvania. 
\end{IEEEbiography}

\begin{IEEEbiography}[{\includegraphics[width=1in,height=1.25in,clip,keepaspectratio]{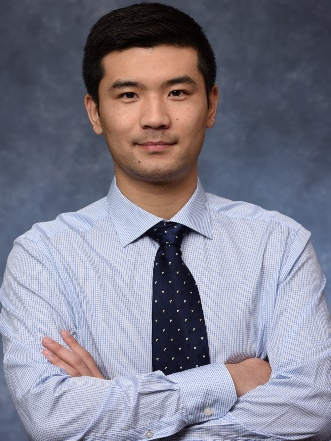}}]{Qingzhou Wan}
received Ph.D degree in Electrical and Computer Engineering from University of Pittsburgh, Pittsburgh, Pennsylvania, in 2022. His research interests include low-dimensional materials based emerging memory, artificial synaptic devices for neuromorphic hardware implementation, and wearable electronics for health monitoring. He is now working at Intel as a TD Mod \& Integr Yield Eng, responsible for developing advanced logic technologies.
\end{IEEEbiography}

\begin{IEEEbiography}[{\includegraphics[width=1in,height=1.25in,clip,keepaspectratio]{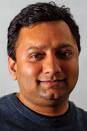}}]{\textbf{Himanshu Akolkar}}
received the M.Tech. degree in electrical engineering from IIT Kanpur, Kanpur, India, and the Ph.D. degree in robotics from IIT Genoa, Genoa, Italy, after which he had a postdoctoral position at Universite Pierre et Marie Curie. He is currently a Postdoctoral Associate at the University of Pittsburgh. His primary research interest includes the neural basis of sensory and motor control to develop intelligent machines.
\end{IEEEbiography}

\begin{IEEEbiography}[{\includegraphics[width=1in,height=1.25in,clip,keepaspectratio]{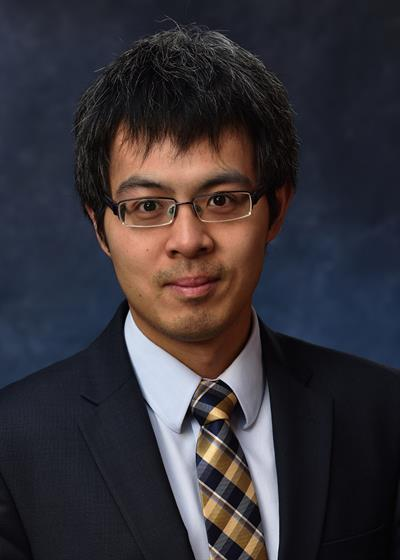}}]{Feng Xiong}
is an Associate professor at the Department of Electrical and Computer Engineering (ECE) at the University of Pittsburgh. Between 2014-2016, he was a postdoctoral fellow at Stanford University. He received his Ph.D. (2014) and M.S. (2010) in ECE from the University of Illinois at Urbana-Champaign (UIUC). He received his B. Eng. with First Class Honor in ECE (2008) from the National University of Singapore (NUS). His research interests include energy-efficient electronics, flexible/wearable electronics, next-generation memory devices (RRAM and PCM), neuromorphic computing,1D and 2D materials, nanoscale thermal transport and renewable energy harvesting. He is the recipient of several awards, including the MRS Graduate Student Gold Award, the TSMC Outstanding Student Research Gold Award, the Ross J. Martin Award, and the Beckman Institute Graduate Fellowship. He is a member of IEEE and MRS.
\end{IEEEbiography}

\begin{IEEEbiography}[{\includegraphics[width=1in,height=1.25in,clip,keepaspectratio]{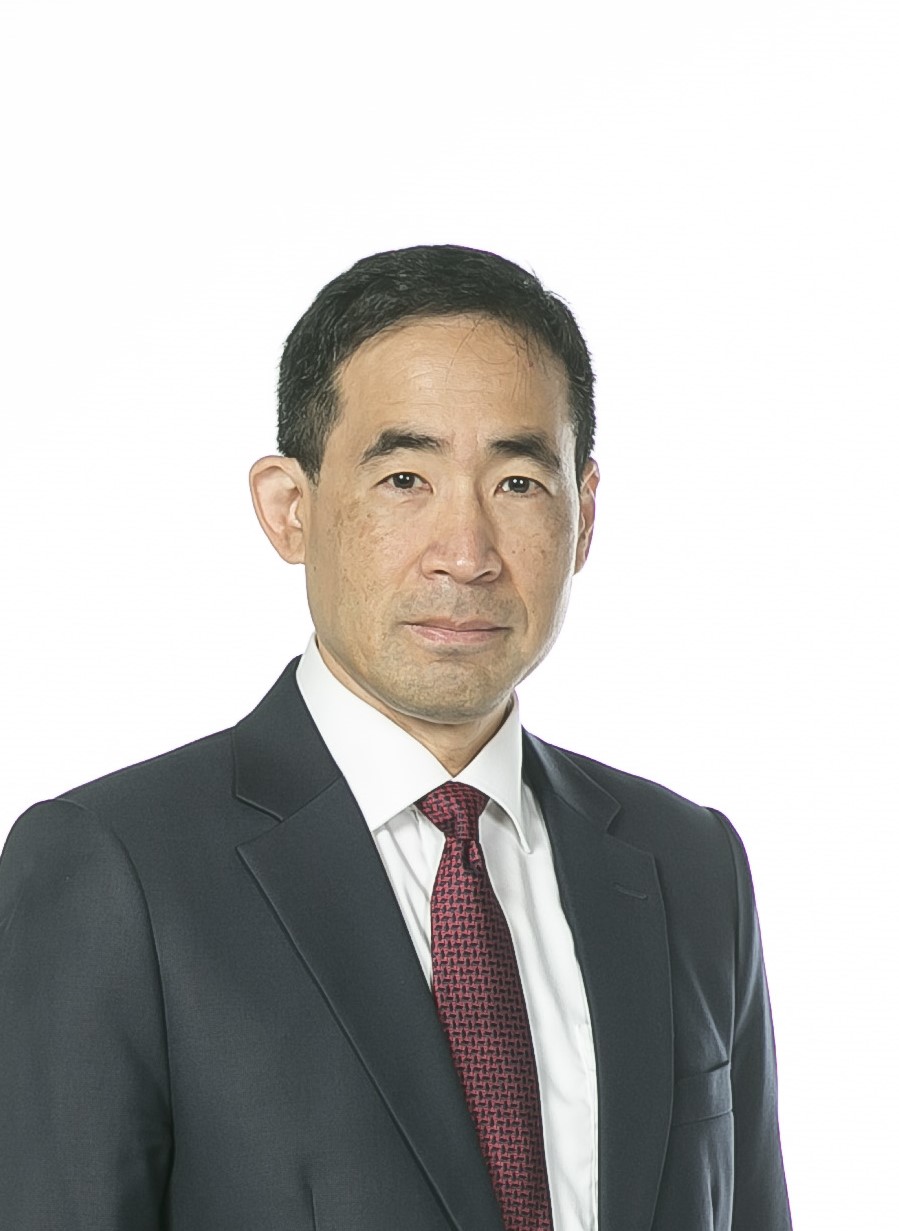}}]{Bertram Shi}
(S'93-M'95-SM'00-F'01) is a Professor of Electronic and Computer Engineering at the Hong Kong University of Science and Technology. His research interests are in bio-inspired signal processing and robotics, neuromorphic engineering, computational neuroscience, developmental robotics, machine vision, image processing, and machine learning.  Prof. Shi twice served as Distinguished Lecturer for the IEEE Circuits and Systems Society.  He has also served as on the editorial boards of the IEEE Transactions on Circuits and Systems, the IEEE Transactions on Biomedical Circuits and Systems and Frontiers in Neuromorphic Engineering.  He served as Chair of the IEEE Circuits and Systems Society Technical Committee on Cellular Neural Networks and Array Computing and as General Chair and Technical Program Chair of conferences in that area.  
\end{IEEEbiography}

\begin{IEEEbiography}[{\includegraphics[width=1in,height=1.25in,clip,keepaspectratio]{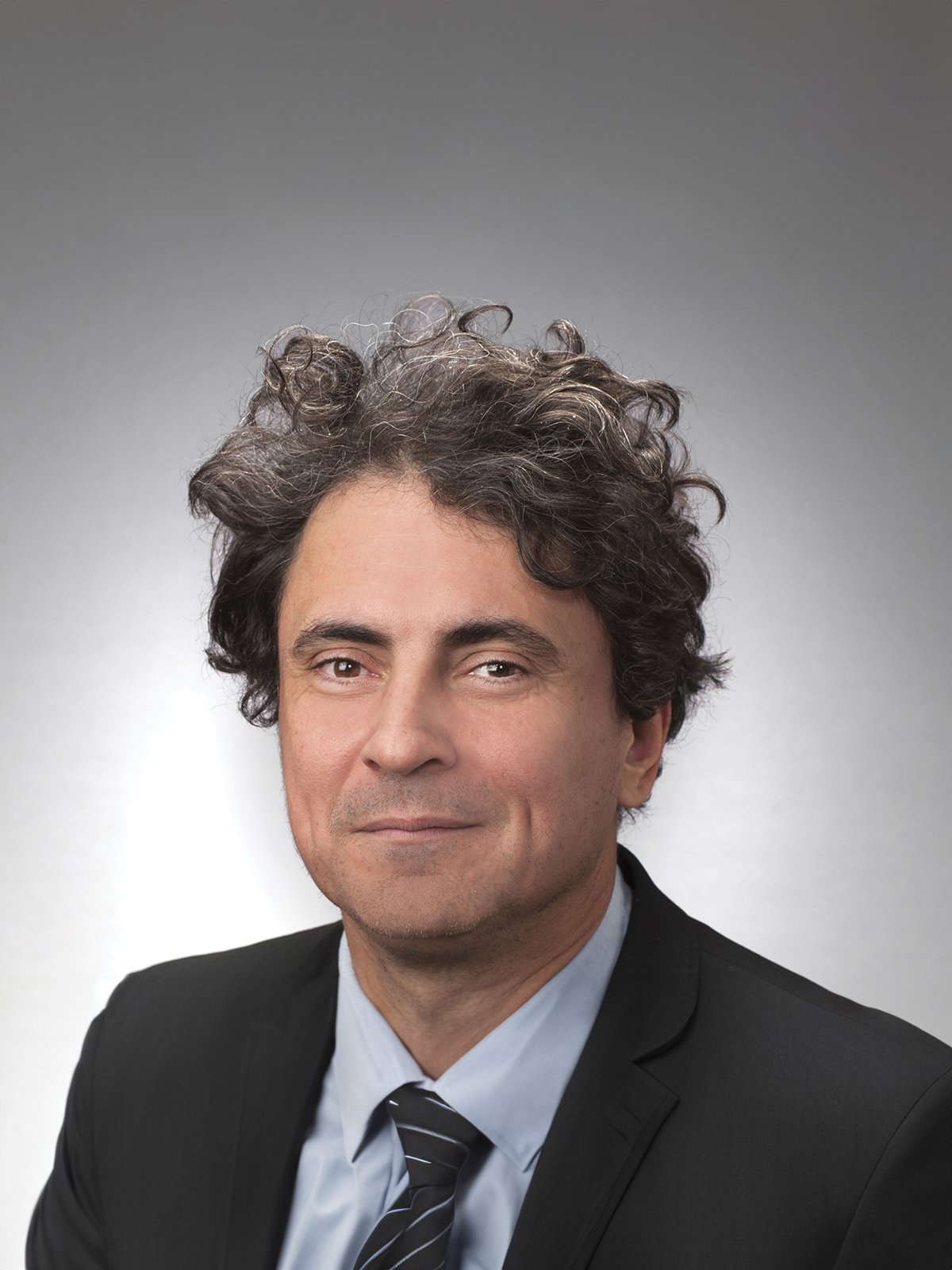}}]{Ryad Benosman}
is a Director of Research at Meta Reality Labs Research, he is adjunct professor at Carnegie Mellon at the Robotics Institute. He received the M.Sc. and Ph.D. degrees in applied mathematics and robotics from the University Pierre and Marie Curie in 1994 and 1999, respectively. He is widely recognized as a pioneer and visionary in neuromorphic event sensing and processing. Dr. Benosman’s primary research goal is to understand the algorithms and mathematics that underlie cortical computation, with the aim of creating new mathematical models and replicating them as functional neuromorphic silicon devices. Dr. Benosman runs a unique research stream that spans from Physiology and primate work to Digital and Analog chip design. This work is seen as a new paradigm of applied neuroscience that merges several traditionally separate fields, such as mathematics, neurosciences, engineering, medicine and hardware design. Dr. Benosman’s work has led to new bioinspired models for AI techniques and sensors that are widely used in academia and industry. He has authored more than 200 peer-reviewed papers and 25 patents, which together form the basis of the field of neuromorphic processing and cognition. Dr. Benosman has launched several companies that utilize his group’s innovations. These include Pixium Vision (retina prosthetics, IPO in 2012), Prophesee (world leading neuromorphic event-based cameras), ThinkLink (motor cortex intention decoding to restore mobility for tetraplegic patients), GraiMatterLabs (event computation and sensor fusion acquired by Snap in 2023).
\end{IEEEbiography}

\newpage

\clearpage

\end{document}